\documentclass[a4paper,10pt]{article}
\usepackage{jheppub}
\usepackage{ifthen}
\usepackage{axodraw}
\usepackage{epsfig}
\usepackage{graphicx}
\usepackage{array}
\usepackage{amsmath,amsthm,amssymb,amsfonts}
\usepackage{array}
\usepackage[center]{subfigure}
\usepackage{dsfont}
\usepackage{mathrsfs}
\usepackage{slashed}
\newcommand{\ba}{\begin{eqnarray}}
\newcommand{\ea}{\end{eqnarray}}
\newcommand{\be}{\begin{equation}}
\newcommand{\ee}{\end{equation}}

\newcommand{\maspace}{\mbox{ }}
\makeatletter
\def\fmslash{\@ifnextchar[{\fmsl@sh}{\fmsl@sh[0mu]}}
\def\fmsl@sh[#1]#2{%
  \mathchoice
    {\@fmsl@sh\displaystyle{#1}{#2}}%
    {\@fmsl@sh\textstyle{#1}{#2}}%
    {\@fmsl@sh\scriptstyle{#1}{#2}}%
    {\@fmsl@sh\scriptscriptstyle{#1}{#2}}}
\def\@fmsl@sh#1#2#3{\m@th\ooalign{$\hfil#1\mkern#2/\hfil$\crcr$#1#3$}}
\makeatother
\title{Renormalisation of heavy-light light ray operators}
\author{M. Knoedlseder,}
\author{N. Offen}
\affiliation{Institut f\"ur theoretische Physik,\\ Universit\"at Regensburg, 93040 Regensburg, Germany}
\emailAdd{michael.knoedlseder@physik.uni-regensburg.de}
\emailAdd{nils.offen@physik.uni-regensburg.de}
\arxivnumber{1105.4569}
\keywords{Heavy Quark Physics, QCD, Conformal and W Symmetry, Renormalization Group}
\abstract{
We calculate the renormalisation of different light ray operators with one light degree of freedom and a static heavy quark. Both
$2\to2$- and $2\to3$-kernels are considered. A comparison with the light-light case suggests that the mixing with three-particle 
operators is solely governed by the light degrees of freedom.
Additionally we show that conformal symmetry is already broken at the level of the one loop counterterms due to the additional 
UV-renormalisation of a cusp in the two contributing Wilson-lines. This general feature can be used to fix the $2\to2$-renormalisation kernels
up to a constant. Some examples for applications of our results are given.}

\begin{document}

\maketitle

%derive the pattern of conformal symmetry breaking and show that for light ray operators including an 
%effective heavy quark 
%$\Gamma_{cusp}$.

\section{Introduction}
Hadrons containing one heavy quark have been one of the most prominent testing grounds of the standard model. 
The elements $\vert V_{ub}\vert$ and $\vert V_{cb}\vert$ of the CKM-matrix are measured in inclusive and exclusive semileptonic decays 
of B-mesons. The angles $\alpha,\,\beta,\,\gamma$ or $\phi_1,\,\phi_2,\,\phi_3$ of the unitarity triangle 
can be measured e.g. in $B\to\pi\pi$, $B\to J/\psi K_s$ or $B\to D K$ decays. Matrix elements of light ray operators are an 
important ingredient of factorisation theorems for these decays.  Brought up by 
\cite{Grozin:1996pq} the phenomenologically most interesting matrix elements of heavy-light light ray operators 
like the B-meson or $\Lambda_b$ distribution amplitudes 
have been under continued scrutiny. Their renormalisation was considered in \cite{Lange:2003ff, Braun:2003wx, Bell:2008er, Ball:2008fw, DescotesGenon:2009hk, Offen:2009mt, Kawamura:2010tj},
while their dependence on the momentum of the light degrees of freedom 
has been analysed either in the sum rule approach \cite{Braun:2003wx, Ball:2008fw, Khodjamirian:2005ea, Khodjamirian:2006st} or in a model 
independent way via operator product expansion \cite{Lee:2005gza, Kawamura:2008vq}. Already in \cite{Braun:2003wx} 
and in the context of inclusive heavy meson decays even earlier in \cite{Grozin:1994ni} it was pointed out 
that, in contrast to the light-light case pioneered by \cite{Efremov:1979qk, Lepage:1980fj}, the special renormalisation 
properties of the heavy-light light ray operators do not allow for an expansion into local operators and therefore no non-negative
moments of the distribution amplitudes can be defined. Despite these efforts unlike to the light-light case 
\cite{Bukhvostov:1985rn, Balitsky:1987bk, Braun:2009vc} no systematic calculation of the renormalisation 
and mixing has been done in the heavy-light case. In this work we try to make the first steps towards a general one loop 
renormalisation of heavy-light light ray operators in coordinate space. We will draw heavily on the results and techniques 
from \cite{Braun:2009vc} although our analysis has the additional problem that one cannot define geometric nor collinear 
twist when an effective heavy quark field is included and therefore at first glance one cannot use the constraints coming 
from conformal invariance of QCD. 
The presentation of our analysis is organised as follows: In section 2 we give some background 
concerning light ray operators, the spinor formalism used in \cite{Braun:2009vc, Braun:2008ia} and conformal symmetry. 
In section 3 we report on the calculation done and give the results for the $2\to2$- and $2\to3$-evolution kernels. 
Section 4 is reserved for the analysis of breaking of conformal symmetry in the renormalisation of heavy-light light ray operators, 
where we will show that even in this case one can derive severe constraints from symmetry arguments. 
In section 5 we will give some examples for applications of our results to the renormalisation of B-meson distribution amplitudes 
and the $\Lambda_b$ distribution amplitude. We conclude in section 6.
\section{Background}
In this section we give a short introduction to some of the theoretical concepts as the spinor representation, 
the definition of light ray operators or conformal symmetry,
while we refrain from giving a detailed account of these topics and rather refer the reader to \cite{Braun:2009vc, Braun:2008ia, Braun:2003rp}.
%introduction to conformal symmetry or the conformal group and instead refer the 
%reader to \cite{Braun:2003rp}.
\subsection{Spinor formalism}
We use the spinor formalism of \cite{Braun:2009vc, Braun:2008ia} solely for classifying the twist of the light 
degrees of freedom and to compare with work in the light-light case since in our explicit calculations it poses 
little advantage due to the absence of Dirac-matrices in the interaction vertices and quark lines of heavy quark effective theory (HQET). 
Therefore we just introduce the basic concepts and refer the reader to \cite{Braun:2009vc, Braun:2008ia} for further details.\\
Via multiplication with the Pauli-matrices 
\[\sigma^\mu=(\mathtt{1},\vec{\sigma}),\qquad\bar{\sigma}^\mu=(\mathtt{1},-\vec{\sigma}),\]
we map each covariant four-vector to a  hermitian matrix $x$:
\begin{equation}
x_{\alpha\dot{\alpha}}=x_\mu(\sigma^\mu)_{\alpha\dot{\alpha}},\qquad \bar{x}_{\dot{\alpha}\alpha}=x_\mu(\bar{\sigma}^\mu)_{\dot{\alpha}\alpha}.
\end{equation}
The Lorentz invariant scalar product can then be expressed via
\begin{equation}
 a_\mu b^\mu=\dfrac{1}{2}a_{\alpha\dot{\alpha}}\bar{b}^{\dot{\alpha}\alpha}=\dfrac{1}{2}\bar{a}^{\dot{\alpha}\alpha}b_{\alpha\dot{\alpha}}
\end{equation}
and Dirac-spinors can be written as two-component Weyl-spinors
\begin{equation}
 q=\left(\begin{array}{c} \psi_\alpha\\\bar{\chi}^{\dot{\beta}}\end{array}\right),\qquad \bar{q}=\left(\chi^\beta,\bar{\psi}_{\dot{\alpha}}\right),
\end{equation}
with $\bar{\psi}_{\dot{\alpha}}=(\psi_\alpha)^\dagger$. The gluon field strength tensor $F_{\mu\nu}$ can be decomposed as follows
\begin{equation}
 F_{\alpha\beta,\dot{\alpha}\dot{\beta}}=\sigma_{\alpha\dot{\alpha}}^\mu\sigma_{\beta\dot{\beta}}^\nu F_{\mu\nu}=2\left(\epsilon_{\dot{\alpha}\dot{\beta}}f_{\alpha\beta}-\epsilon_{\alpha\beta}\bar{f}_{\dot{\alpha}\dot{\beta}}\right),
\end{equation}
where $\epsilon_{\alpha\beta}$ is the two-dimensional antisymmetric Levi-Civita tensor and $f_{\alpha\beta}$, $\bar{f}_{\dot{\alpha}\dot{\beta}}$
are chiral and antichiral (or self-dual and anti-self-dual) symmetric tensors which belong to the $(1,0)$ and $(0,1)$ representations of the Lorentz-group. 
Their explicit expression can be written, taking the covariant derivative $D_\mu=\partial_\mu-igA_\mu$, as
\begin{equation}
 f_{\alpha\beta}=\dfrac{1}{4}\left(D_\alpha^{\maspace\dot{\alpha}}\bar{A}_{\dot{\alpha}\beta}+D_\beta^{\maspace\dot{\alpha}}\bar{A}_{\dot{\alpha}\alpha}\right),\quad \bar{f}_{\dot{\alpha}\dot{\beta}}=\dfrac{1}{4}\left(\bar{D}_{\dot{\alpha}}^{\maspace\alpha}A_{\alpha\dot{\beta}}+\bar{D}_{\dot{\beta}}^{\maspace\alpha}A_{\alpha\dot{\alpha}}\right),
\end{equation}
For going over from the Dirac to the spinor representation the following relations come in handy
\begin{equation}
 f_{\alpha\beta}=\dfrac{i}{4}\sigma^{\mu\nu}_{\alpha\beta}F_{\mu\nu},\qquad \bar{f}_{\dot{\alpha}\dot{\beta}}=-\dfrac{i}{4}\bar{\sigma}_{\dot{\alpha}\dot{\beta}}^{\mu\nu}F_{\mu\nu},
\end{equation}
where $\sigma_{\mu\nu}$ is expressed via the Pauli-matrices $\sigma_\mu$
\begin{equation}
 \left(\sigma^{\mu\nu}\right)_\alpha^{\maspace\beta}=\dfrac{i}{2}\left[\sigma^\mu\bar{\sigma}^\nu-\sigma^\nu\bar{\sigma}^\mu\right]_\alpha^{\maspace\beta},\quad\left(\bar{\sigma}^{\mu\nu}\right)^{\dot{\alpha}}_{\maspace\dot{\beta}}=\dfrac{i}{2}\left[\bar{\sigma}^\mu\sigma^\nu-\bar{\sigma}^\nu\sigma^\mu\right]^{\dot{\alpha}}_{\maspace\dot{\beta}}
\end{equation}
and the expressions for Dirac-matrices in the spinor basis are given by:
\begin{equation}
 \gamma^\mu=\left(\begin{array}{cc} 0&\left[\sigma^\mu\right]_{\alpha\dot{\beta}}\\\left[\bar{\sigma}^\mu\right]^{\dot{\alpha}\beta}&0\end{array}\right),\quad\fmslash{n}=\left(\begin{array}{cc} 0 & n_{\alpha\dot{\beta}}\\\bar{n}^{\dot{\alpha}\beta}&0\end{array}\right),
\end{equation}
\begin{equation}
 \sigma^{\mu\nu}=\left(\begin{array}{cc} \left[\sigma^{\mu\nu}\right]_\alpha^{\maspace\beta}&0\\0&\left[\bar{\sigma}^{\mu\nu}\right]^{\dot{\alpha}}_{\maspace\dot{\beta}}\end{array}\right),\quad \gamma_5=\left(\begin{array}{cc} -\delta_\alpha^\beta&0\\0&\delta_{\dot{\beta}}^{\dot{\alpha}}\end{array}\right).
\end{equation}
To define plus and minus components we introduce two light-like vectors which in general can be represented 
as a product of two spinors which we denote $\lambda$ and $\mu$
\begin{eqnarray}
 n_{\alpha\dot{\alpha}}&=&\lambda_\alpha\bar{\lambda}_{\dot{\alpha}},\quad n^2=0,\nonumber\\
\tilde{n}_{\alpha\dot{\alpha}}&=&\mu_\alpha\bar{\mu}_{\dot{\alpha}},\quad \tilde{n}^2=0,
\end{eqnarray}
with $\bar{\lambda}=\lambda^*$ and $\bar{\mu}=\mu^*$. Arbitrary four-vectors can be decomposed into components along 
and transverse to the light rays
\begin{equation}
 x_{\alpha\dot{\alpha}}=z\lambda_\alpha \bar{\lambda}_{\dot{\alpha}}+\tilde{z}\mu_\alpha \bar{\mu}_{\dot{\alpha}}+w\lambda_\alpha\bar{\mu}_{\dot{\alpha}}+\bar{w}\mu_\alpha\bar{\lambda}_{\dot{\alpha}},\quad x^2=(\mu\lambda)(\bar{\lambda}\bar{\mu})\left[z\tilde{z}-w\bar{w},\right]
\end{equation}
where $z,\,\tilde{z}$ and $w,\bar{w}=w^*$ are the real respective complex coordinates in the two light-like directions and 
the transverse plane. Finally, the $+$ and $-$ components of the fields are defined as projections onto $\lambda$ and $\mu$ spinors
\begin{equation}
 \begin{array}{lll} \psi_+=\lambda^\alpha\psi_\alpha, & \chi_+=\lambda^\alpha \chi_\alpha,& f_{++}=\lambda^\alpha\lambda^\beta f_{\alpha\beta},\\
  \bar{\psi}_+=\bar{\lambda}^{\dot{\alpha}}\bar{\psi}_{\dot{\alpha}},&\bar{\chi}_+=\bar{\lambda}^{\dot{\alpha}}\chi_{\dot{\alpha}},&\bar{f}_{++}=\bar{\lambda}^{\dot{\alpha}}\bar{\lambda}^{\dot{\beta}}\bar{f}_{\dot{\alpha}\dot{\beta}},\\
\psi_-=\mu^\alpha \psi_\alpha,&\bar{\psi}_-=\bar{\mu}^{\dot{\alpha}}\bar{\psi}_{\dot{\alpha}},&f_{+-}=\lambda^\alpha\mu^\beta f_{\alpha\beta},\end{array}
 \end{equation}
with quantum numbers under the special conformal group as in table \ref{tab:class}.
\begin{table}[h]
\begin{center}
 \begin{tabular}{|c|c|c|c|c|c|c|}
  \hline\hline
&$\psi_+$&$f_{++}$&$\psi_-$&$f_{+-}$&$\bar{D}_{-+}\psi_+$&$\bar{D}_{-+}f_{++}$\\\hline
$j$&1&3/2&1/2&1&3/2&2\\\hline
$E$&1&1&2&2&2&2\\\hline
$H$&1/2&1&-1/2&0&3/2&2\\\hline
 \end{tabular}
\caption{Conformal spin $j$, twist $E$ and helicity $H$ of the primary fields taken from \cite{Braun:2009vc}.\label{tab:class}}
\end{center}
\end{table}

\subsection{Heavy-light light ray operators}
%We follow \cite{Balitsky:1987bk, Braun:2009vc} in defining light ray operators as non-local gauge-invariant operators whose fields ly on a light-like line $n^2=0$. Here we will deal with heavy-light light ray operators which means that one of the fields is an effective heavy quark field. 
Perhaps the most well-known example of a heavy-light light ray operator is the one whose matrix element between 
a B-meson state and the vacuum defines the B-meson distribution amplitude $\phi_B^+$ \cite{Grozin:1996pq} which 
is a main ingredient in most factorisation theorems for exclusive B-decays. It can be written as a product of a 
light and a heavy quark field at light-like distance
\begin{equation}
 \mathcal{O}(z_1,z_2)=\bar{q}(z_1 n)\fmslash{n}[z_1,z_2]h_v(z_2 n)
\label{eq:op1}
\end{equation}
where $n^\mu$ is a light-like vector $n^2=0$ and $[z_1,z_2]$ is the path-ordered exponential
\begin{equation}
 [z_1,z_2]=P\exp\left\{i g z_{12}\int_0^1 d\alpha\, n^\mu A_\mu(z_{12}^\alpha n)\right\}.
\end{equation}
Here and throughout the paper we use the short-hand notations
\begin{equation}
 z_{12}=z_1-z_2,\qquad z_{12}^\alpha=\bar{\alpha} z_1+\alpha z_2,\qquad \bar{\alpha}=1-\alpha 
\end{equation}
and we will write $\Phi(z_1)$ instead of $\Phi(z_1 n)$ for a field living on the light cone in order not to overburden our formulae.\\
The scale dependence of (\ref{eq:op1}) is governed by the renormalisation group equation
\begin{equation}
 \left(\mu\dfrac{\partial}{\partial \mu}+\beta(g)\dfrac{\partial}{\partial g}+\dfrac{\alpha_s}{2\pi}\mathcal{H}\right)\left[\mathcal{O}(z_1,z_2)\right]_R=0,
\end{equation}
where $\beta(g)$ is the QCD beta function and $\mathcal{H}$ is the integral operator \cite{Lange:2003ff,Braun:2003wx, Kawamura:2010tj}
%was first worked out in \cite{Lange:2003ff} in momentum space and in \cite{Braun:2003wx} see also \cite{Kawamura:2010tj} in position space.
\begin{equation}
 \left[\mathcal{H} \mathcal{O}_1\right](z_1,z_2)=2C_F\left[\int_0^1\dfrac{d\alpha}{\alpha}\left(\mathcal{O}(z_1,z_2)-\bar{\alpha}\mathcal{O}(z_{12}^\alpha,z_2)\right)+\log(i\mu \,z_{12})-\dfrac{5}{4}\right].
\end{equation}
For the purpose of this paper we follow \cite{Braun:2009vc} in defining light ray fields $\Phi$ as fields living on the light cone multiplied by a Wilson-line
\begin{equation}
\Phi(z)=[0,z]\Phi(z),\qquad \bar{\Phi}(z)=\bar{\Phi}(z)[z,0],\label{eq:def-field}
\end{equation}
where the Wilson-line has to be taken in the proper representation depending on whether $\Phi$ is a gluon or a quark field. A gauge invariant heavy-light light ray operator is then nothing else than a product of light ray fields with a proper invariant colour tensor $S$ where at least one of the fields is an effective heavy quark field and one a light quark or gluon:
\begin{equation}
 \mathcal{O}(z_1,\ldots,z_N)=S\left(\Phi(z_1)\otimes\ldots\otimes\Phi(z_N) \right).
\end{equation}
Our analysis considers operators composed out of the following fields
\begin{equation}
 \Phi=\left\{h_v,\,\psi_+,\,\psi_-,\,f_{++},\,f_{+-}\right\}
\end{equation}
and their respective complex conjugates. %or antichiral fields.
Taking the fields $\chi_+,\,\chi_-$ instead of $\psi_+,\,\psi_-$ makes no change whatsoever. Their classification 
with respect to conformal spin, twist and helicity is the same and can be found together with those of the other 
light degrees of freedom in table \ref{tab:class}. Though we are aware of the fact that due to the heavy quark field 
our operators are in no representation of the conformal group we use the same notation as in \cite{Braun:2009vc}. 
The $\otimes$ just indicates that the fields have open colour indices so that
\begin{equation}
(t^a\otimes t^a) (\Phi(z_1)\otimes\Phi(z_2))=t^a\Phi(z_1)\otimes t^a\Phi(z_2)=t^a_{i_1i'_1}\Phi^{i'_1}(z_1) t^a_{i_2i'_2}\Phi^{i'_2}(z_2),
\end{equation}
where the generators of the $SU(3)$ have to be taken in the appropriate representation:
\begin{eqnarray}
 (t^a\psi)^i&=&t^a_{ii'}\psi^{i'},\quad\;\;\;(t^a\,h_v)^i=t^a_{ii'}h_v^{i'},\nonumber\\
(t^a\bar{\psi})^i&=&-t^a_{i'i}\bar{\psi}^{i'},\quad(t^a\,h^*_v)^i=-t^a_{i'i}\,h_v^{*i'},\nonumber\\
(t^a f)^b&=&if^{bab'}f^{b'}.
\end{eqnarray}
%for mutual comparison.

%We follow \cite{Braun:2009vc} in defining light ray fields as fields living on the light cone $n^2=0$ multiplied by a Wilson-line in the appropriate representation, e.g:
%\begin{equation}
% h_v(z)\to[0,z]h_v(z),\qquad \psi_+(z)\to [0,z]\psi_+(z),\qquad f_{++}(z)\to [0,z]f_{++}(z).
%\end{equation}
%where $\psi_+,\,f_{++}$ are projections on the plus components of the quark field and gluon-field-strength-tensor. The generators in the Wilson-lines are given by: 
%\begin{eqnarray}
% (t^a\psi(h_v))^i&=&t^a_{ii'}\psi^{i'}(h_v^{i'}),\nonumber\\
%(t^a\bar{\psi}(h_v^*))^i&=&-t^a_{i'i}\bar{\psi}^{i'}(h_v^{*i'}),\nonumber\\
%(t^a f)^b&=&if^{bab'}f^{b'}.
%\end{eqnarray}

\subsection{Renormalisation group equations and light cone gauge}
Since operators with the same quantum numbers can mix under renormalisation, the renormalisation constant $Z$
\begin{equation}
 \left[\mathcal{O}_i(\Phi)\right]_R=Z_{ik}\mathcal{O}_k(\Phi_0),
\end{equation}
with $\Phi_0=Z^{1/2}_\Phi\Phi$ and therefore the anomalous dimension
\begin{equation}
 \gamma=-\mu\dfrac{d}{d\mu}\mathbb{Z}\mathbb{Z}^{-1},\qquad \gamma=\dfrac{\alpha_s}{2\pi}\mathbb{H},
\label{eq:def-kernels}
\end{equation}
entering the renormalisation group equation
\begin{equation}
 \left(\mu\dfrac{\partial}{\partial \mu}+\beta(g)\dfrac{\partial}{\partial g}+\gamma_{ik}\right)\left[\mathcal{O}_k(\Phi)\right]_R=0,
\end{equation}
are $n\times n$-matrices if $\mathcal{O}_i(\Phi)$ with $i=1,\ldots,n$ is a complete set of operators closed under 
renormalisation.  $\gamma$ and $\mathbb{H}$ have block triangular form since $N$-particle operators can only mix with 
operators with $M\geq N$ particles not the other way round. At one loop level it can be shown that the diagonal elements 
of the anomalous dimension matrix are given by sums of $2\to2$-kernels which is seen most explicitly in the light cone gauge 
$n\cdot A=0$ that we use throughout our calculations. In this gauge the Wilson-lines are just identity matrices and therefore 
the relevant one loop diagrams reduce to simple exchange diagrams. Similarly the off-diagonal elements reduce to sums of 
$2\to3$-kernels. Take for example the operator
\begin{equation}
 \mathcal{O}_3=S\left(h_v\otimes f_{++}\otimes \bar{\psi}_+\right)
\end{equation}
where $S=t^a_{ij}$, related to the combination of three-particle distribution amplitudes $\tilde{\Psi}_A-\tilde{\Psi}_V$ of the B-meson. Its renormalisation can be built out of the kernels for the operators
\begin{equation}
 h_v\otimes f_{++},\quad h_v\otimes \bar{\psi}_+,\quad f_{++}\otimes \bar{\psi}_+,
\end{equation}
which can be written as
\begin{eqnarray}
 \mathbb{H}\, h^i_v(z_1) f^a_{++}(z_2)&=& -2\left(t_{ii'}^b \,t_{aa'}^b\right)\left[\mathcal{H}_h-\sigma_g-\sigma_h\right]h_v^{i'}(z_1) f_{++}^{a'}(z_2),\nonumber\\
\mathbb{H}\, h^i_v(z_1) \bar{\psi}^j_+(z_2)&=& -2\left(t_{ii'}^b\, t_{jj'}^b\right)\left[\mathcal{H}_h-\sigma_q-\sigma_h\right]h_v^{i'}(z_1) \bar{\psi}_+^{j'}(z_2),\nonumber\\
\mathbb{H}\, f^a_{++}(z_1) \bar{\psi}^i_+(z_2)&=&-2\left(t_{aa'}^b\, t_{ii'}^b\right)\left[\hat{\mathcal{H}}-2\mathcal{H}^+-\sigma_q-\sigma_g\right]f_{++}^{a'}(z_1) \bar{\psi}_+^{i'}(z_2)\nonumber\\
&&+4\left(t^{a'}t^a\right)_{ii'}\mathcal{H}^-f_{++}^{a'}(z_1)\bar{\psi}_+^{i'}(z_2).
\end{eqnarray}
Explicit expressions for the heavy-light kernel $\mathcal{H}_h$ will be given in the next section while $\hat{\mathcal{H}},\,\mathcal{H}^+$ and $\mathcal{H}^-$ can be found in \cite{Braun:2009vc}.\\
A drawback of using the light cone gauge is the explicit breaking of Lorentz invariance so that plus and minus components of the fields renormalise differently \cite{Bassetto:1987sw}
\begin{equation}
 \left[q_\pm\right]_0=Z_\pm^{1/2} q_\pm,\qquad \left[A_\mu\right]_0=R_\mu^{\maspace\nu} A_\nu,
\label{eq:ren-konst1}
\end{equation}
where $Z_\pm$ and $R_\mu^{\maspace\nu}$ are at one loop given by
\begin{equation}
 Z_+=1+\dfrac{3\alpha_s}{4\pi\varepsilon}C_F,\qquad Z_-=1-\dfrac{\alpha_s}{4\pi\varepsilon}C_F
\end{equation}
and
\begin{equation}
 R_{\mu\nu}=Z_3^{1/2}\left[g_{\mu\nu}-\left(1-\tilde{Z}_3^{-1}\right)\dfrac{n_\mu\bar{n}_\nu}{n\cdot \bar{n}}\right],
\label{eq:ren-gluon}
\end{equation}
with
\begin{equation}
 Z_3=1+\dfrac{\alpha_s}{4\pi\varepsilon}\left(\dfrac{11}{3}N_c-\dfrac{2}{3}n_f\right),\quad \tilde{Z}_3=1+\dfrac{\alpha_s}{2\pi\varepsilon}N_c.
\end{equation}
After explicit calculation we need only three constants $\sigma_g,\,\sigma_q,\,\sigma_h$ which are defined as in \cite{Braun:2009vc} and in Appendix A, Eq. (\ref{eq:sigmah})
\begin{eqnarray}
 \sigma_q&=&\dfrac{3}{4},\qquad \sigma_h=\dfrac{1}{2},\nonumber\\
\sigma_g&=&\dfrac{b_0}{4N_c},\qquad b_0=\dfrac{11}{3} N_c-\dfrac{2}{3} n_f,
\end{eqnarray}
with
\begin{equation}
 Z_q^{1/2}=1+\dfrac{\alpha_s}{2\pi\varepsilon} \sigma_q C_F,\quad Z_g^{1/2}=1+\dfrac{\alpha_s}{2\pi\varepsilon}\sigma_g C_A,\quad Z_h^{1/2}=1+\dfrac{\alpha_s}{2\pi\varepsilon}\sigma_h C_F,
\end{equation}
to take into account the renormalisation of the fields and the constant terms appearing in the $2\to2$-kernels. 

%A general feature that can be shown is, that the counter terms to the product of fields $\Phi\otimes\Phi$ is again given b
%which are obviously not gauge invariant but according to the definition \ref{eq:def-field} transform like the product of two gauge matrices $U(0)\otimes U(0)$ at space time point $x=0$ in the appropriate representation.

\subsection{Conformal invariance and conformal group \label{sec:conf}}
Massless QCD is at classical level conformally invariant. This property is broken at one loop level by the conformal anomaly 
\cite{Adler:1976zt, Collins:1976yq, Nielsen:1977sy, Minkowski:1976en} but nevertheless can be used to constrain one loop counterterms 
since these are essentially tree-level objects.\\
The conformal group is the largest generalisation of the Poincar$\acute{\mbox{e}}$-group that leaves the light cone invariant. 
Conformal transformations include in addition to Lorentz-rotations and translations, dilatations, inversions and special conformal 
transformations:
\begin{eqnarray}
 x^\mu&\longrightarrow&x'^\mu=\lambda x^\mu,\nonumber\\
 x^\mu&\longrightarrow&x'^\mu=\dfrac{x^\mu}{x^2},\nonumber\\
 x^\mu&\longrightarrow&x'^\mu=\dfrac{x^\mu+a^\mu x^2}{1+2a\cdot x+a^2 x^2}.
\end{eqnarray}
The full conformal algebra consists of fifteen generators, where ten, translations $P^\mu$ and Lorentz-rotations $M^{\mu\nu}$, 
come from the Poincar$\acute{\mbox{e}}$-group, one from dilatations, $D$, and four from the special conformal transformations, 
$K^\mu$. These generators act on a generic fundamental field $\Phi$ with arbitrary spin as
\begin{eqnarray}
 \delta_P^\mu\,\Phi(x)&=&i\left[\mathbf{P}^\mu,\Phi(x)\right]=\partial^\mu\Phi(x),\nonumber\\
\delta_M^{\mu\nu}\,\Phi(x)&=&i\left[\mathbf{M}^{\mu\nu},\Phi(x)\right]=\left(x^\mu\partial^\nu-x^\nu\partial^\mu-\Sigma^{\mu\nu}\right)\Phi(x),\nonumber\\
\delta_D\,\Phi(x)&=&i\left[\mathbf{D},\Phi(x)\right]=\left(x\cdot \partial +l\right)\Phi(x),\nonumber\\
\delta_k^\mu\,\Phi(x)&=&i\left[\mathbf{K}^\mu,\Phi(x)\right]=\left(2x^\mu x\cdot\partial-x^2\partial^\mu+2lx^\mu-2x_\nu\Sigma^{\mu\nu}\right)\Phi(x).
\label{eq:diff1}
\end{eqnarray}
Here $l$ is the canonical dimension of the field and $\Sigma^{\mu\nu}$ the generator of spin rotations:
\[\Sigma^{\mu\nu}\phi(x)=0,\quad \Sigma^{\mu\nu}\psi(x)=\dfrac{i}{2}\sigma^{\mu\nu}\psi(x),\quad\Sigma^{\mu\nu}A^\alpha(x)=g^{\nu\alpha}A^\mu(x)-g^{\mu\alpha}A^\nu(x),\]
where $\phi$, $\psi$ and $A^\alpha$ are a scalar- a fermion- and a vector-field.
Of special interest for fields living on a light ray $\Phi(z)=\Phi(z n)$ is the collinear subgroup $SL(2,\mathbb{R})$ which 
generates projective transformations on a line. It's generators are habitually written in the following form:
\begin{equation}
 \begin{array}{ll}
  \mathbf{S}_+=-i\mathbf{P}_+,& \mathbf{S}_0=\frac{i}{2}(\mathbf{D}+\mathbf{M}_{-+}),\\
\mathbf{S}_-=\frac{i}{2}\mathbf{K}_-,&\mathbf{E}\:\,=\frac{i}{2}(\mathbf{D}-\mathbf{M}_{-+}).
 \end{array}
\label{eq:diff-coll}
\end{equation}
Their action on quantum fields can be similarly to Eq. (\ref{eq:diff1}) written as differential operators acting on 
the field coordinates
\begin{equation}
 S_-=z^2 \partial_z+2jz,\quad S_0=z\partial_z+j,\quad S_+=-\partial_z,
\end{equation}
while $\mathbf{E}$ commutes with all $\mathbf{S}_i$ and counts the twist of the fields
\[\left[\mathbf{E},\Phi(x)\right]=\dfrac{1}{2}(l-s)\Phi(x)=E\Phi(x),\]
where $l$ is again the canonical dimension, $s$ is the spin projection along the light ray and $j$ is the conformal spin defined as
\[j=\dfrac{1}{2}(l+s)=l-\dfrac{E}{2}.\]
Conformal symmetry even if it is anomalous implies that the one loop renormalisation kernels $\mathbb{H}$, see Eq. 
(\ref{eq:def-kernels}), commute with the generators of the conformal group. For fields living on the light ray
this condition reduces to the generators of the collinear subgroup.\\ 
We will state some of the basics that are needed for our analysis in section 4.
By Noethers theorem every symmetry induces a conserved current. For the dilatation and special conformal transformation 
these currents are given by
\begin{equation}
 J_D^\mu=x_\nu \Theta^{\mu\nu},\qquad J^\mu_{K\alpha}=(2x_\nu x_\alpha-x^2g_{\nu\alpha})\Theta^{\mu\nu},
\end{equation}
where $\Theta^{\mu\nu}$ is the modified, symmetric and traceless, energy momentum tensor of QCD \cite{Callan:1970ze}. Obviously 
these currents are conserved on the classical level but quantum corrections introduce a scale and therefore violate dilatation invariance.
This so called trace anomaly is given by \cite{Adler:1976zt, Collins:1976yq, Nielsen:1977sy, Minkowski:1976en}
\begin{eqnarray}
 \partial_\mu J_D^\mu(x)&=&g_{\mu\nu} \Theta^{\mu\nu}(x)\stackrel{EOM}{=}\Delta_D(x)\stackrel{EOM}{=}\dfrac{\beta(g)}{2g} G_{\mu\nu}^a G^{a\mu\nu}(x)\nonumber\\
&=&(D-4)\dfrac{1}{4} G_{\mu\nu}^a G^{a\mu\nu}(x)+\mathcal{O}(\alpha_s),
\label{eq:conform-anom}
\end{eqnarray}
where $\Delta_D$ is defined as
\begin{equation}
 \Delta_D=(l+1)\dfrac{\partial \mathcal{L}}{\partial(\partial_\mu\Phi)}\partial_\mu\Phi+l\dfrac{\partial\mathcal{L}}{\partial\Phi}-D\mathcal{L}
\end{equation}
and determines the variation of the action under dilatation
\begin{equation}
\delta_\alpha S=\alpha\, \delta_D S=\alpha\int d^Dx \,\Delta(x),\qquad\Delta(x)=\Delta_D(x)-(D-2)\partial^\lambda \mathcal{O}_{B\lambda}(x)
\label{eq:delta_D}
\end{equation}
where $\alpha$ is an infinitesimal parameter and $ \mathcal{O}_{B\lambda}$ a BRST-exact operator \cite{Belitsky:1998vj, Belitsky:1998gc, Braun:2003rp}
which plays only a minor role in our forthcoming analysis. For a definition and some details see Appendix B.
$EOM$ means that we are dealing with classical solutions of the equations of motion. For the current of special conformal 
transformations an additional factor of $2x^\nu$ appears:
\begin{equation}
 \delta_\alpha=\alpha_\nu\, \delta^\nu_K S=\alpha_\nu\int d^Dx\,2x^\nu \,\Delta(x).
\label{eq:delta_K}
\end{equation}
%We have omitted a BRST-exact operator that contributes to the conformal anomaly \cite{Belitsky:1998vj, Belitsky:1998gc, Braun:2003rp} 
%since it does not play any role in our forthcoming calculations, for some details see Appendix B. 
The scale invariance of the renormalised action is therefore broken by terms of 
$\mathcal{O}(\alpha_s)$ or by terms proportional to $D-4$, where $D$ is the number of space-time dimensions. 
In \cite{Derkachov:2010zza} a simple proof is given that the one loop counterterms nevertheless exhibit conformal symmetry 
and we will rely heavily on their work in section 4. There it will be seen that if an effective heavy quark field participates, 
the aforementioned statement no longer holds and that this fact can be traced back to the additional UV-renormalisation 
of the cusp of two Wilson-lines \cite{Korchemsky:1987wg}.

%The fact that the variation of the action is proportional to $D-4$ is the reason why

\section{Calculation and results}
This section gives a short account of the calculation and the relevant results, showing that only one $2\to2$-renormalisation 
kernel $\mathcal{H}_h$ governs the evolution of all heavy-light light ray operators and that the $2\to3$-mixing coincides with 
the light-light case if the effective heavy quark is substituted by a chiral plus component of a light quark, e.g. $\psi_+$.
\subsection{Calculation}
Throughout our calculation we used light cone gauge $n\cdot A=0$ or $A_{11}=0$. This eliminates the Wilson-lines 
associated with the light ray fields but gives an additional term in the gluon-propagator:
\begin{eqnarray}
 d^{ab}_{\mu\nu}(q)&=&-i\dfrac{\delta^{ab}}{q^2+i\epsilon}\left(g_{\mu\nu}-\dfrac{q_\mu n_\nu+q_\nu n_\mu}{n\cdot q}+q_\mu q_\nu\dfrac{n^2+\xi q^2}{(n\cdot q)^2}\right)\nonumber\\
&=&-i\dfrac{\delta^{ab}}{q^2+i\epsilon}\left(g_{\mu\nu}-\dfrac{q_\mu n_\nu+q_\nu n_\mu}{n\cdot q}\right),\qquad n^2=0,\quad \xi=0,
\label{eq:gluon-prop}
\end{eqnarray}
with $\xi$ being the gauge parameter.
We habitually get rid of the spurious pole in the second term by using
\begin{equation}
  \dfrac{e^{iq_+(z_1-z_2)}}{q_+}=i(z_1-z_2)\int_0^1d\alpha\, e^{i\alpha q_+(z_1-z_2)}+ \dfrac{1}{q_+}, 
\end{equation}
with $q_+=n\cdot q$, where the second term gives just a local in most cases divergent constant and we use the Mandelstam-Leibbrandt 
prescription \cite{Mandelstam:1982cb, Leibbrandt:1983pj}
\begin{equation}
\dfrac{1}{n\cdot q}\longrightarrow \dfrac{\tilde{n}\cdot q}{n\cdot q\;\tilde{n}\cdot q +i\epsilon}
\label{eq:Mand-Leib}
\end{equation}
for its explicit calculation. Lets consider as an example the easiest case $h_v\otimes\psi_+$:
Since the chirality of the light quark does not matter for the $2\to2$-kernel we can for simplicity just use 
$\bar{q}(z_2)\fmslash{n} h_v(z_1)$ and calculate its matrix element $M$ with on-shell quarks to one loop order (see figure \ref{fig:2-to-2})
\begin{equation}
M^{(1)}= -iC_F g_s^2\left(\dfrac{\mu}{2\pi}\right)^{4-D}\int \dfrac{d^Dl}{(2\pi)^D}\dfrac{1}{l_+\,l^2}\left[\dfrac{1}{v\cdot l}-2\dfrac{k_+-l_+}{(k-l)^2}\right]e^{il_+(z_2-z_1)}e^{-ik_+ z_2} \bar{v}(k)\fmslash{n}u(v),
\end{equation}
which after above procedure gives:
\begin{eqnarray}
 M^{(1)}&=&(z_1-z_2)C_F g_s^2\left(\dfrac{\mu}{2\pi}\right)^{4-D}\int \dfrac{d^Dl}{(2\pi)^D}\int_0^1d\alpha\dfrac{1}{l^2}\left[\dfrac{1}{v\cdot l}-2\dfrac{k_+-l_+}{(k-l)^2}\right]\nonumber\\
&\times&e^{i\alpha l_+(z_2-z_1)}e^{-ik_+ z_2} \bar{v}(k)\fmslash{n}u(v)\nonumber\\
&-&iC_F g_s^2\int\dfrac{d^Dl}{(2\pi)^4}\dfrac{1}{l_+\,l^2}\left[\dfrac{1}{v\cdot l}-2\dfrac{k_+-l_+}{(k-l)^2}\right]e^{-ik_+ z_2} \bar{v}(k)\fmslash{n}u(v).
\label{eq:res-calc1}
\end{eqnarray}
The first two terms are the same one would get in Feynman-gauge while those in the third row are due to the additional term in (\ref{eq:gluon-prop}).
Calculating the integrals one gets
\begin{eqnarray}
 M^{(1)}&=&C_F\dfrac{\alpha_s}{2\pi\varepsilon}\left[\int_0^1d\alpha \dfrac{\bar{\alpha}}{\alpha}\left(1-e^{i\alpha k_+(z_2-z_1)}\right)+\dfrac{1}{2\varepsilon}+\log(i\mu(z_2-z_1))\right.\nonumber\\
&&-\left.1+\int_0^\infty \dfrac{dl_+}{l_+}\left(\dfrac{\mu}{l_+}\right)^{2\varepsilon}\right]\bar{v}(k_+)\fmslash{n}u(v)e^{-ik_+z_2},
\label{eq:res-calc2}
\end{eqnarray}
where one sees the same structure as in \cite{Braun:2003wx, Kawamura:2010tj} and in the second row one term that cancels the 
difference between the renormalisation constants of the light quark in Feynman- and light cone gauge and an additional scaleless integral
which if one regularises it to get only the ultra-violet divergence is canceled by the same integral appearing in the renormalisation 
of the heavy quark in light cone gauge, see Appendix A.\\
Similarly a little care has to be taken in the case of $h_v\otimes f_{+-}$ since $f_{+-}$ includes transverse as well as minus components
of the gluon field (and two transverse gluon fields) which are renormalised differently, see (\ref{eq:ren-gluon}). 
In the $2\to2$-kernel for $h_v\otimes f_{+-}$ there appears an additional term proportional to $h_v\otimes A_{--}$ which exactly cancels the 
difference in renormalisation of $A_{--}$ and $A_{+-}$ so that one does not have to introduce a new constant for $f_{+-}$ 
in (\ref{eq:kern2}). As shown below the $\log(i\mu(z_2-z_1))$- and $\frac{1}{\varepsilon^2}$-term are a general feature of the 
renormalisation of heavy-light light ray operators related to $\Gamma_{cusp}$ and as already pointed out in \cite{Braun:2003wx} 
and \cite{Grozin:1994ni} it is exactly this $\log$-term that hinders the expansion into local operators because it is 
obviously singular for $z_2-z_1=0$.\\
Ignoring the last term and taking the derivative of (\ref{eq:res-calc2}) with respect to $\log\mu$ one gets the evolution kernel for the operator 
$h_v\, \psi_+$ which will among others be given in the next section.
\subsection{$2\to2$-kernels}
\begin{figure}[h]
 \begin{center}
  \epsfig{file=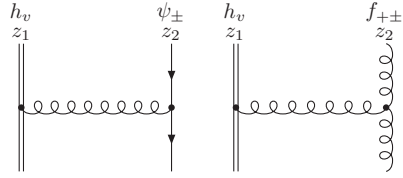, scale=0.8}
\caption{Diagrams contributing to the $2\to2$-kernels in light cone gauge. \label{fig:2-to-2}}
 \end{center}
\end{figure}
For the heavy-light $2\to2$-kernels only the diagrams shown in figure \ref{fig:2-to-2} contribute nontrivially and there appears
a single function $\mathcal{H}_h$ which depends solely on the conformal spin $j$ of the light degree of freedom:
\begin{equation}
 \left[\mathcal{H}_h \mathcal{O}\right](z_1,z_2)=\int_0^1\dfrac{d\alpha}{\alpha}\left(\mathcal{O}(z_1,z_2)-\bar{\alpha}^{2j-1}\mathcal{O}(z_1,z_{21}^\alpha)\right)+\log(i\mu (z_2-z_1)).
\label{eq:2kernels}
\end{equation}
The results do not, except for a sign, depend on the chirality of the light degrees of freedom nor if one considers a $\psi$- or a $\chi$-spinor. 
They are given simply by multiplying $\mathcal{H}_h$ with the appropriate colour structure and adding the respective constants $\sigma_q,\,\sigma_g$ and $\sigma_h$:
\begin{eqnarray}
 \mathbb{H}\, h^i_v(z_1) f^a_{++}(z_2)&=& -2\left(t_{ii'}^b\, t_{aa'}^b\right)\left[\mathcal{H}_h-\sigma_h-\sigma_g\right]h_v^{i'}(z_1) f_{++}^{a'}(z_2),\label{eq:kern1}\\
\mathbb{H}\, h^i_v(z_1) f^a_{+-}(z_2)&=& -2\left(t_{ii'}^b\, t_{aa'}^b\right)\left[\mathcal{H}_h-\sigma_h-\sigma_g\right]h_v^{i'}(z_1) f_{+-}^{a'}(z_2),\label{eq:kern2}\\
\mathbb{H}\, h^i_v(z_1) \psi^j_+(z_2)&=& -2\left(t_{ii'}^b\,t_{jj'}^b\right)\left[\mathcal{H}_h-\sigma_h-\sigma_q\right]h_v^{i'}(z_1) \psi_+^{j'}(z_2),\label{eq:kern3}\\
\mathbb{H}\, h^i_v(z_1)\psi^j_-(z_2)&=& -2\left(t_{ii'}^b\, t_{jj'}^b\right)\left[\mathcal{H}_h-\sigma_h-\sigma_q\right]h_v^{i'}(z_1) \psi_-^{j'}(z_2)\label{eq:kern4}.
\end{eqnarray}
The form of (\ref{eq:2kernels}) does not come unexpected. The first part resembles the contribution coming from light 
degrees of freedom seen in \cite{Braun:2009vc} as well and the heavy quark just gives a contribution coming from the 
renormalisation of two intersecting Wilson-lines, 
one light-like and one time-like.\footnote{The heavy quark can be written as a sterile quark field $\phi(-\infty)$ multiplied by a 
time-like Wilson-line \cite{Korchemsky:1991zp} \[h_v(z_1)=P\exp\left\{ig_s\int_{-\infty}^{0}d\alpha\, v^\mu A_\mu(\alpha v+z_1 n)\right\}\phi(-\infty)\].} 
\subsection{$2\to3$-kernels}
\begin{figure}[h]
\begin{center}
 \epsfig{file=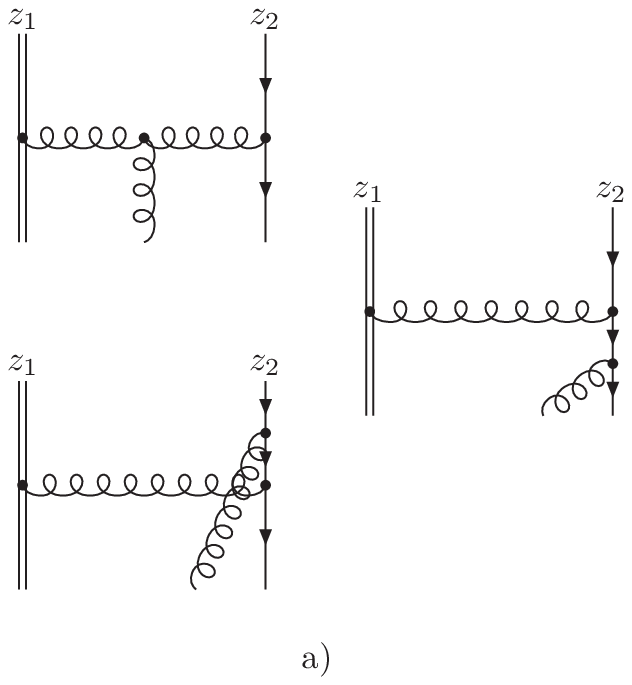, scale=0.6}\hspace{1.5cm}\epsfig{file=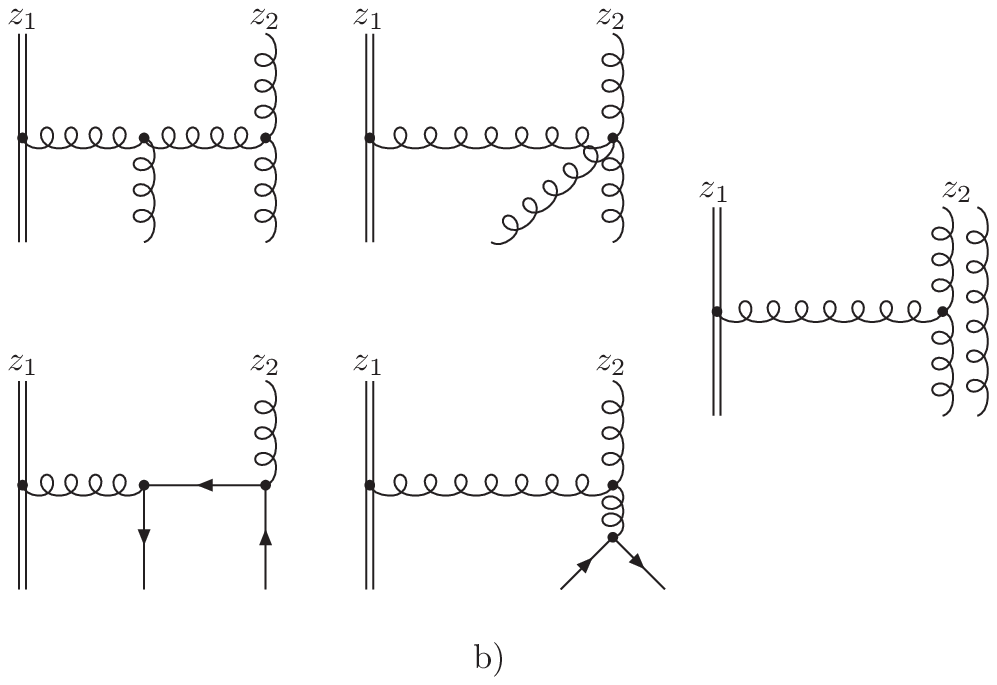, scale=0.6}
\caption{ Wilson-lines are omitted since light cone gauge $n\cdot A=0$ is used. a) Single diagram contributing to kernels (\ref{eq:kern5}). b) Diagrams contributing to kernels (\ref{eq:kern6})
\label{fig:diag-kern}} 
\end{center}
\end{figure}

For the two $2\to3$-kernels we calculated, we closely follow the notation of \cite{Braun:2009vc}. Before we give the results
let us recall some of the abbreviations used there.\\
In light cone gauge the one loop renormalisation of an operator $[X]_R(z_1,z_2)$ can in general be written as
\begin{equation}
\left[X\right]_R(z_1,z_2)=X_B(z_1,z_2)+\dfrac{\alpha_s}{4\pi\varepsilon}\left[\mathbb{H}^{(2\to2)}X\right](z_1,z_2)+\dfrac{\alpha_s}{4\pi\varepsilon}\left[\mathbb{H}^{(2\to3)}Y\right](z_1,z_2)
\end{equation}
where $X_B(z_1,z_2)$ is the bare operator and the relevant $2\to2$-kernels $\mathbb{H}^{(2\to2)}$ have been given in the preceding section, 
though we would have to add the $\frac{1}{\varepsilon^2}$-poles here. 
We first consider the simpler case $X^{ij}(z_1,z_2)=h_v^i(z_1)\psi^j_-(z_2)$. It mixes with just a single operator 
\[Y^{ija}=g(\mu\lambda) h_v^i(z_1)\psi_+^j(z_2)\bar{f}_{++}^a(z_3)\]
 and there are only two different colour structures for the three-particle counter term
\begin{equation}
 \left[\mathbb{H}^{(2\to3)}Y\right]^{ij}=\left\{f^{abc}t^b_{ii'}t^c_{jj'}\mathcal{H}_1+it^b_{ii'}(t^dt^b)_{jj'}\mathcal{H}_2\right\}Y^{i'j'a},
\label{eq:color-13}
\end{equation}
which in light cone gauge follows from the Feynman-diagrams shown in figure \ref{fig:diag-kern} a).\\
For $X^{ia}(z_1,z_2)=h^i_v(z_1) f^a_{+-}(z_2)$ the calculation is more complicated, there are five diagrams 
(see figure \ref{fig:diag-kern} b)) contributing, it mixes with three different operators
\begin{eqnarray}
 Y^{iad}(z_1,z_2,z_3)&=&g(\mu\lambda)h^i_v(z_1)f_{++}^a(z_2)\bar{f}_{++}^d(z_3),\nonumber\\
J^{ia}(z_1,z_2)&=&g(\mu\lambda)h^i_v(z_1)\left(\bar{\psi}_+(z_2) t^a \psi_+(z_2)+\chi_+(z_2) t^a\bar{\chi}_+(z_2)\right),\nonumber\\
Z^{iab}(z_1,z_2,z_3)&=&g(\mu\lambda)h_v^i(z_1)\left(\bar{\psi}_+(z_2) t^a t^b\psi_+(z_3)-\chi_+(z_2) t^b t^a \bar{\chi}_+(z_3)\right),\nonumber\\
\label{eq:op-23}
\end{eqnarray}
and there appear four different colour structures
\begin{eqnarray}
\left[\mathbb{H}^{(2\to3)}Y\right]^{ia}&=&\Big(\left\{f^{dbc}t^b_{ii'}f^{aca'}\mathcal{H}_1+i(t^dt^b)_{ii'}f^{aba'}\mathcal{H}_2\right\}Y^{i'a'd}\nonumber\\
&+& t_{ii'}^bf^{aba'}\tilde{\mathcal{H}}_1 J^{i'a'}+t_{ii'}^b\tilde{\mathcal{H}}_2 Z^{i'ab}\Big).
\label{eq:color-23}
\end{eqnarray}
The results are given below where a comparison with the results for the counterterms of $\psi_+\otimes\psi_-$ and $f_{+-}\otimes\psi_+$ 
from \cite{Braun:2009vc} shows that they coincide if one substitutes $h_v\to \psi_+$. This indicates that the $2\to3$-mixing is 
solely governed by the twist of the light degrees of freedom. In addition we could not find any extraordinary mixing under renormalisation due to the heavy quark.\\
\begin{enumerate}

%\noindent
%\textbf{\underline{1. $X^{ij}(z_1,z_2)=h_v^i(z_1)\psi_-^j(z_2)$}}\\
\item  $X^{ij}(z_1,z_2)=h_v^i(z_1)\psi_-^j(z_2)$.\\
As stated there is only a single operator $Y^{ijd}(z_1,z_2,z_3)=g(\mu\lambda)h^i_v(z_1) \psi^j_+(z_2)\bar{f}^d_{++}(z_3)$ 
needed in this case. The two kernels appearing in (\ref{eq:color-13}) are given by:
\begin{eqnarray}
\left[\mathcal{H}_1 Y\right](z_1,z_2)&=&z_{12}^2\int_0^1d\alpha\int_0^{\bar{\alpha}} d\beta\,\beta\,Y(z_1,z_{21}^\alpha,z_{12}^\beta),\nonumber\\
\left[\mathcal{H}_2 Y\right](z_1,z_2)&=&z_{12}^2\int_0^1d\alpha\int_{\bar{\alpha}}^1d\beta\dfrac{\bar{\alpha}\bar{\beta}}{\alpha}\,Y(z_1,z_{21}^\alpha,z_{12}^\beta).
\label{eq:kern5}
\end{eqnarray}
%\noindent
%\textbf{\underline{2. $X^{ia}(z_1,z_2)=h_v^i(z_1) f^a_{+-}(z_2)$:}}\\
\item  $X^{ia}(z_1,z_2)=h_v^i(z_1) f^a_{+-}(z_2)$.\\
Here the three operators from (\ref{eq:op-23}) contribute with colour structures and kernels specified as in (\ref{eq:color-23}). 
The latter can be written as 
\begin{eqnarray}
\left[\mathcal{H}_1 Y\right](z_1,z_2)&=&z_{12}^2\int_0^1d\alpha\int_0^{\bar{\alpha}}d\beta\,\bar{\alpha}\beta\,Y(z_1,z_{21}^\alpha,z_{12}^\beta),\nonumber\\
\left[\mathcal{H}_2 Y\right](z_1,z_2)&=&z_{12}^2\int_0^1d\alpha\int_{\bar{\alpha}}^1 d\beta\dfrac{\bar{\alpha}^2\bar{\beta}}{\alpha}\,Y(z_1,z_{21}^\alpha,z_{12}^\beta),\nonumber\\
\left[\tilde{\mathcal{H}}_1 J\right](z_1,z_2)&=&-z_{12}\int_0^1d\alpha\,\bar{\alpha}^2\,J(z_1,z_{21}^\alpha),\nonumber\\
\left[\tilde{\mathcal{H}}_2 Z\right](z_1,z_2)&=&z_{12}\int_0^1d\alpha\int_{\bar{\alpha}}^1 d\beta\dfrac{\bar{\beta}}{\beta}\,Z(z_1,z_{12}^\alpha,z_{21}^\beta),
\label{eq:kern6}
\end{eqnarray}
where for brevities sake we omitted all colour indices.
\end{enumerate}

\section{Constraints from conformal symmetry and\\ breaking of scale invariance due to the heavy quark}
As was seen in section 3 the $2\to2$-kernels are functions of only one variable $z_2-z_1$ so for simplicity we use here $z_1=0$
and $z_2=t$.\footnote{For $z_1\neq0$ we would have to take $z_1$ as the centre of the inversions in $v\cdot K$}
To use constraints from conformal symmetry we first have to determine the behaviour of the heavy quarks 
under conformal transformations. The heavy quark can be written as a time-like Wilson-line times a sterile scalar field \cite{Korchemsky:1991zp}:
\begin{equation}
 h_v(0)=P\exp\left\{ig_s\int_{-\infty}^{0}d\alpha\, v^\mu A_\mu(\alpha v)\right\}\phi(-\infty).
\end{equation}
There are two conformal transformations that map the respective time-like Wilson-line onto itself: the special conformal transformation along the 
$v$-direction where $v_\mu=\frac{1}{2}(n_\mu+\tilde{n}_\mu)$ and the dilatation, for definitions see section \ref{sec:conf}. The generators of 
these transformations are given by %\footnote{A time-like vector $v^\mu$ or a light-like vector $n^\mu$ are both transformed 
%into a time-like or a light-like vector}:
\begin{eqnarray}
 i\left[v\cdot \mathbf{K},\Phi(x)\right]&=&\left[2v\cdot x x\cdot \partial-x^2v\cdot \partial +2 lv\cdot x-2v_\mu x_\nu \Sigma^{\mu\nu}\right]\Phi(x)=v\cdot K\,\Phi(x),\nonumber\\
i\left[\mathbf{D},\Phi(x)\right]&=&\left[x\cdot \partial +l\right]\Phi(x)=D\,\Phi(x),
\label{eq:conf-trafos}
\end{eqnarray}
with
\[i\left[v\cdot\mathbf{K},\mathbf{D}\right]=-v\cdot \mathbf{K}.\]
As introduced in section 2.4, $\Phi$ is either a scalar, a spinor or a vector field, $l$ is the canonical dimension of the field and $\Sigma^{\mu\nu}$ is the 
generator of spin rotations. %As in \cite{Braun:2003rp} we use boldface letters to 
% which has the form
%\begin{equation}
% \Sigma^{\mu\nu}\phi(x)=0,\qquad \Sigma^{\mu\nu}\psi(x)=\dfrac{i}{2}\sigma^{\mu\nu}\psi(x),\qquad\Sigma^{\mu\nu}A^\alpha=g^{\nu\alpha}A^\mu-g^{\mu\alpha}A^\nu.
%\label{eq:spin-gener}
%\end{equation}
For fields living on the light cone 
\[\Phi(z)=\Phi(z n),\quad n^2=0\] 
the two generators take on an especially simple form:
\begin{eqnarray}
i\left[v\cdot \mathbf{K},\Phi(z)\right]&=&(2z^2 \partial_z+4 j z)\Phi(z)=v\cdot K\,\Phi(z),\label{eq:generators}\\
i\left[\mathbf{D},\Phi(z)\right]&=&(z\partial_z+l)\Phi(z)=D\,\Phi(z),
\label{eq:generators2}
\end{eqnarray}
where $j=\frac{1}{2}(l+s)$ with $\Sigma_{+-}\Phi(z)=s\Phi(z)$ is the conformal spin of the light field. In particular $v\cdot\mathbf{K}$ 
is reduced to $\tilde{n}\cdot\mathbf{K}$ since a special conformal transformation in the $n$-direction has no effect altogether. Additionally there is no 
interchange of plus and minus components of the fields under $v\cdot\mathbf{K}$ since such terms would be proportional to transverse coordinates. 
This can be seen explicitly if one writes down the generator of special conformal transformations in spinor notation, $v\cdot \mathbf{K}=\mathbf{K}_{1\dot{1}}+\mathbf{K}_{2\dot{2}}$:
\begin{eqnarray}
 i\left[\mathbf{K}_{\alpha\dot{\alpha}},\psi_\beta\right](x)&=&
\left(x_{\alpha\dot{\gamma}}x_{\gamma\dot{\alpha}}\partial^{\gamma\dot{\gamma}}+4x_{\alpha\dot{\alpha}}\right)\psi_\beta(x)-2x_{\beta\dot{\alpha}}\psi_\alpha,\\
i\left[\mathbf{K}_{\alpha\dot{\alpha}},f_{\beta\delta}\right](x)&=&
\left(x_{\alpha\dot{\gamma}}x_{\gamma\dot{\alpha}}\partial^{\gamma\dot{\gamma}}+6x_{\alpha\dot{\alpha}}\right)f_{\beta\delta}(x)-2x_{\beta\dot{\alpha}}f_{\alpha\delta}-2x_{\delta\dot{\alpha}}f_{\alpha\beta}.
\end{eqnarray}
Heavy-light light ray operators therefore behave in a well defined way under this transformations. What happens after renormalisation? 
From the explicit form of the renormalisation kernels (\ref{eq:2kernels}) and the differential operators (\ref{eq:generators}), (\ref{eq:generators2}) 
the following commutation relations for $\mathcal{H}$ with $D$ and $v\cdot K$ are derived:
\begin{eqnarray}
 \left[D,\mathcal{H}_h\right]\mathcal{O}(t)&=&\mathcal{O}(t),\label{eq:invar1a}\\
\left[v\cdot K ,\mathcal{H}_h\right]\mathcal{O}(t)&=&0.
\label{eq:invar1b}
\end{eqnarray}
%They show that scale invariance is broken in general and invariance under special conformal transformations by the position of the heavy-quark. Taking a closer look at the calculation it is seen that the breaking of scale invariance is due to the $\log$-term in $\mathcal{H}_h$ and therefore due to the cusp-anomalous dimension of a light-like and a time-like Wilson-line. The same is true for the invariance under special conformal transformation but this is not seen directly from this calculation. There is another way in which these properties can be shown even a priori i.e. before calculating the renormalisation kernels. 
%They show a remnant of the conformal symmetry of the light-light case: Special conformal transformations leave the Hamiltonians $\mathcal{H}$ invariant and the dilatation-invariance is only broken by the $\log$ term. These results could have been obtained a priori by following \cite{Derkachov:2010zza}. There it was shown that one loop counter terms inherit the symmetry under scale and conformal transformations from the action which therefore are symmetries of the Hamiltonian $\mathcal{H}$. 
\begin{figure}[h]
\begin{center}
\epsfig{file=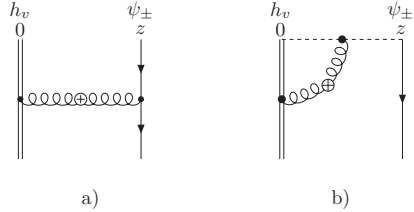, scale=0.75}
\caption{Feynman-diagrams which are responsible for the breaking of the conformal invariance in a) light cone gauge b) Feynman-gauge. 
A change of the light degrees of freedom does not alter the calculation substantially in light cone gauge. 
The cross denotes the insertion of $(D-4)\int d^Dx\, G_{\mu\rho}^aG^{a\mu\rho}(x)$ or  $2(D-4)\int d^Dx\,x^\nu G_{\mu\rho}^aG^{a\mu\rho}(x)$ . \label{fig:anom}}
\end{center}
\end{figure}
They show that the variation of the operator under dilatation and its renormalisation do not commute which means that in contrast 
to the case of pure massless QCD, where scale invariance is broken at one loop order only by finite terms and therefore the renormalisation 
of light-light light ray operators is not affected to this order, the inclusion of an effective heavy quark gives a contribution that breaks scale invariance 
already at the level of the one loop counterterms.
We will proceed to show that this phenomenon can be directly traced back to the cusp at $z=0$ in the time-like Wilson-line representing
the effective heavy quark and the light-like Wilson-lines included for gauge invariance. 
In \cite{Derkachov:2010zza} a simple proof was given that the one loop counterterms inherit conformal symmetry from the 
Lagrange-density. We will apply their results to the case at hand. 
Let $\mathcal{O}(\Phi)$ be a two-particle operator with an effective heavy quark and $\Delta \mathcal{O}(\Phi)$ be its counterterm. 
Therefore Green-functions with an insertion of $\mathcal{O}(\Phi)+\Delta\mathcal{O}(\Phi)$ or equivalent the path integral
\begin{equation}
G_{\mathcal{O}}(\lambda)=\int [D\Phi]\exp\left\{-S_R+i\int d^4x\,\lambda\,\Phi\right\}(\mathcal{O}(\Phi)+\Delta\mathcal{O}(\Phi))
\end{equation}
are finite. Here $S_R$ denotes the renormalised action of QCD and HQET while $\Phi$ is any one of the relevant fields 
$A_\mu,\,\psi,\,h_v,\ldots$ and $\lambda$ the respective source. Making a change of variables 
$\Phi\to\Phi'=\Phi+\delta_\alpha\Phi$ where $\delta_\alpha=\alpha\delta_D\Phi=\alpha\, i[\mathbf{D},\Phi]$ does 
not change the path integral and leads to the relation:
\begin{align}
 &\int [D\Phi]\exp\left\{iS_R+i\int d^4x\,\lambda\,\Phi\right\}(\delta_\alpha\mathcal{O}(\Phi)+\delta_\alpha\Delta\mathcal{O}(\Phi))\\
&=\int[D\Phi]\exp\left\{iS_R+i\int d^4x\,\lambda\,\Phi\right\}\left(\delta_\alpha S_R-i\int d^4x\lambda\delta_\alpha\Phi\right)\times(\mathcal{O}(\Phi)+\Delta\mathcal{O}(\Phi)).
\label{eq:invar2}
\end{align}
As stated, equation (\ref{eq:invar1a}) implies that the counterterm of the variation of the operator is 
not identical to the varied counterterm
\begin{equation}
 \Delta(\delta_\alpha\mathcal{O}(\Phi))\neq\delta_\alpha(\Delta\mathcal{O}(\Phi)),
\label{eq:counterterm-form}
\end{equation}
for $\delta_\alpha=\alpha\delta_D$ and we proceed to show that this follows directly from the right hand side of equation (\ref{eq:invar2}). 
The term proportional to the sources $\lambda$ is finite at one loop order \cite{Derkachov:2010zza}, therefore the only relevant term is
\begin{equation}
\int[D\Phi]\exp\left\{iS_R+i\int d^4x\,\lambda\,\Phi\right\}\delta_\alpha S_R\times\mathcal{O}(\Phi).
\label{eq:insert}
\end{equation}
Using that the variation under (global) dilatation of the action is given by (\ref{eq:conform-anom}), (\ref{eq:delta_D}), see also \cite{Braun:2003rp}
\begin{eqnarray}
 \delta_{D}S&=&\int d^Dx \Delta_D(x),\nonumber\\
&=&\int d^Dx (D-4)G_{\mu\nu}^aG^{a\mu\nu}(x)+\mathcal{O}(\alpha_s),
\label{eq:insert2}
%\qquad \delta_{S_+}S=2\epsilon^\alpha\int d^Dx x_\alpha \Delta(x)
\end{eqnarray}
%with
%\begin{equation}
% \Delta(x)=\Delta_D(x)-(D-2)\partial^\lambda\mathcal{O}_{B\lambda}(x)
%\end{equation}
%where $\mathcal{O}_{B\lambda}$ is a BRST-exact operator, see \cite{Braun:2003rp} that is irrelevant for our discussion and that the trace anomaly $\Delta_{S_0}$ has the form \cite{Adler:1976zt, Collins:1976yq, Nielsen:1977sy, Minkowski:1976en}
%\begin{equation}
% \Delta_D(x)=(D-4) G_{\mu\nu}^aG^{a\mu\nu}(x)+\mathcal{O}(\alpha_s)
%\label{eq:insert2}
%\end{equation}
where we omitted the total derivative of $\mathcal{O}_{B\lambda}$, one can calculate the integral (\ref{eq:insert}) at one loop level. In light cone gauge there is just the 
exchange diagram with an insertion of (\ref{eq:insert2}). One needs an $\frac{1}{\varepsilon^2}$-pole so that after the 
multiplication with $D-4$ from (\ref{eq:insert2}) there remains a divergent part which would proof (\ref{eq:counterterm-form}). 
The relevant contribution comes from the additional term in the gluon propagators and that part of (\ref{eq:insert2}) where the 
derivatives in the field-strength tensor cancel one of the denominators. 
For the example in figure \ref{fig:anom} a) these contributions to (\ref{eq:insert}) amount to
\begin{equation}
 \delta_\alpha=\alpha\,\delta_D:\quad C_Fg_s^2(D-4)\int \dfrac{d^Dl}{(2\pi)^D}\dfrac{1}{l_+\,v\cdot l\,(k-l)^2}e^{-i(k_+-l_+)z}%e^{-il_+z_1},\nonumber\\
% \delta_\alpha=\delta_{S_+}:&&2 z_1\,C_F g_s^2\epsilon\int \dfrac{d^Dl}{(2\pi)^D}\dfrac{1}{l_+\,v\cdot l\,(k-l)^2}e^{-i(k_+-l_+)z_2}e^{-il_+z_1}
\label{eq:invar3} 
\end{equation}
which after integration confirms (\ref{eq:invar1a}). The calculation does not substantially depend on the light degree of freedom which is explicitly seen
if one does the same derivation in Feynman-gauge. Analysing the relevant diagrams only those of figure \ref{fig:anom} b) contribute
which clearly shows that only the colour structure of the result depends on the light degree of freedom and as anticipated
the breaking of scale invariance comes due to the additional UV-renormalisation from the cusp in the
two Wilson-lines \cite{Korchemsky:1987wg}, where one is light-like included for gauge-invariance and the other is time-like and is representing 
the effective heavy quark. An explicit calculation again reproduces the results in equation (\ref{eq:invar1a}) 
though the colour structures only match for gauge invariant operators. It should be noted that these $\frac{1}{\varepsilon^2}$-poles appear 
only for the two-particle counterterms while the three-particle terms are unaffected. A fact that supports the assumption that the 
mixing with three-particle operators is constrained by the transformation properties with respect to the conformal group of the light 
degrees of freedom and which probably could be further exploited.\\
The same computation as for the dilatation can be done for the case of $v\cdot\mathbf{K}$ 
but here the additional $x^\nu$ in $\delta_K S$, see (\ref{eq:delta_K}), gives an extra propagator denominator, 
so that the relevant integral amounts to
\begin{equation}
 \delta_\alpha=\alpha\,\delta_{v\cdot K}:\quad 2C_Fg_s^2(D-4)\int \dfrac{d^Dl}{(2\pi)^D}\dfrac{1}{l_+}\left[\dfrac{1}{(v\cdot l)^2\,(k-l)^2}+\dfrac{1}{(k-l)^4}\right]e^{-i(k_+-l_+)z},%e^{-il_+z_1},\nonumber\\
% \delta_\alpha=\delta_{S_+}:&&2 z_1\,C_F g_s^2\epsilon\int \dfrac{d^Dl}{(2\pi)^D}\dfrac{1}{l_+\,v\cdot l\,(k-l)^2}e^{-i(k_+-l_+)z_2}e^{-il_+z_1}
\label{eq:invar_K} 
\end{equation}
which gives only a finite contribution and in this way confirms (\ref{eq:invar1b}).\footnote{A subtlety in covariant gauges is related 
to the variation of the gauge-fixing term which gives a divergent contribution. This only vanishes for gauge invariant operators, 
see Appendix B.}\\
Another interesting point following from this considerations is that the $2\to2$ evolution kernels are fixed up to a constant by the 
constraints (\ref{eq:invar1a}) and (\ref{eq:invar1b}). It is not possible to construct a finite integral kernel $\mathcal{H}$ of a single 
variable which commutes with both $D$ and $v\cdot K$ except for a constant and therefore it is not possible to write
down an evolution kernel that fulfils the constraints (\ref{eq:invar1a}) and (\ref{eq:invar1b}) that differs from (\ref{eq:2kernels}) by more 
than said constant. This can be understood in the following way: An integral kernel of a single variable that is invariant 
under dilatations has the form
\begin{equation}
 \left[\mathcal{H}\mathcal{O}\right](z)=A\int_0^z\dfrac{dt}{z}\,f\left(\dfrac{t}{z}\right)\,\mathcal{O}(t)+B\mathcal{O}(z),
\end{equation}
where $f$ is a generic function that gives a finite integral and $A$ and $B$ are arbitrary constants. This can be written in the more familiar way
\begin{equation}
\left[\mathcal{H}\mathcal{O}\right](z)=A\int_0^1d\alpha\,f\left(\bar{\alpha}\right)\,\mathcal{O}(\bar{\alpha}z)+B\mathcal{O}(z).
\label{eq:dilatation-con}
\end{equation}
Now taking the differential operator for $v\cdot K$ (\ref{eq:generators}) and calculating the commutator $[v\cdot K,\mathcal{H}]\mathcal{O}(t)$ 
one gets the following differential equation for $f(\bar{\alpha})$:
\begin{equation}
 \alpha\bar{\alpha}\dfrac{d}{d\alpha}f(\bar{\alpha})+(1-2\alpha(1-j))f(\bar{\alpha})=0,
\label{eq:diff-eq}
\end{equation}
with the boundary conditions
\[\lim_{\alpha\to1}\alpha\bar{\alpha}f(\bar{\alpha})=0,\qquad \lim_{\alpha\to0}\alpha\bar{\alpha}f(\bar{\alpha})=0.\]
The solution to (\ref{eq:diff-eq})
\begin{equation}
 f(\bar{\alpha})=C\dfrac{\bar{\alpha}^{2j-1}}{\alpha}
\label{eq:solution1}
\end{equation}
violates the boundary conditions and therefore only a constant can commute with both $D$ and $v\cdot K$.\footnote{To make the integral 
\[\int_0^1d\alpha\,f(\bar{\alpha})\mathcal{O}(\bar{\alpha}z)\] well defined, one should regularise it by writing $f(\bar{\alpha})$ e.g. as 
a $+$-distribution.} Taking this argument a little further one can use that (\ref{eq:invar3}) and therefore (\ref{eq:invar1a}) and (\ref{eq:invar1b}) 
are a general feature of the renormalisation of heavy-light operators and then construct $\mathcal{H}_h$ up to a constant by exploiting these constraints.\footnote{The colour 
structure of (\ref{eq:invar1a}) and (\ref{eq:invar1b}) depends on the gauge if the operator is not gauge invariant.}
In light cone gauge the term violating scale invariance has the same colour factor as the rest of the renormalisation kernel. We add
a $\log (i\mu z)$-term to (\ref{eq:dilatation-con}) which then is the most general expression that fulfils (\ref{eq:invar1a}). 
Solving the constraint (\ref{eq:invar1b}) then amounts to solving (\ref{eq:diff-eq}) with the changed boundary condition
% king light cone gauge one can simply add a $\log (i\mu z)$-term with the same colour factor as the integral term to (\ref{eq:dilatation-con}) 
%violating dilatation invariance in the proper way and then solve 
%the equation resulting from $v\cdot K$.
%This amounts to solving (\ref{eq:diff-eq}) with the changed boundary condition
\[\lim_{\alpha\to0}\alpha\bar{\alpha}f(\bar{\alpha})=-1,\]
where (\ref{eq:solution1}) is now a viable and therefore unique solution, except for a constant. With the regularised $f(\bar{\alpha})$
\[f(\bar{\alpha})=\left(-\dfrac{\bar{\alpha}^{2j-1}}{\alpha}\right)_+,\]
where
\[\int_0^1d\alpha \left(-\dfrac{\bar{\alpha}^{2j-1}}{\alpha}\right)_+\mathcal{O}(\bar{\alpha} z)=\int_0^1d\alpha\dfrac{\bar{\alpha}^{2j-1}}{\alpha}(\mathcal{O}(z)-\mathcal{O}(\bar{\alpha}z))\]
we then get (\ref{eq:2kernels}) apart for an unconstrained constant.
%though it has to be kept in mind that their colour structure is gauge dependent  

%In this way we are able to predict that the renormalisation kernels for all heavy-light $2\to2$-kernels 
%have the form (\ref{eq:2kernels}).

%and gives a direct link between the breaking of conformal invariance in the renormalisation of heavy-light 
%light ray operators and the cusp anomalous dimension.
%RECHNUNG EXPLIZITER AUFSCHREIBEN UND ERGEBNIS VIA GLEICHUNG (34) VERDEUTLICHEN! ZUDEM ANFUEHREN BZW. ZEIGEN, DASS
%DIE HERGELEITETEN BEDINGUNGEN AN DIE EVOLUTIONSKERNE DIESE EINDEUTIG FESTLEGEN. (SO MOEGLICH!)
%ZUDEM ANBRINGEN, DASS SICH DIR RECHNUNG AUCH IN LICHTKEGELEICHUNG DURCHFUEHREN LIESSE, DORT ABER WENIGER INSTRUKTIV IST.
\section{Applications}
Matrix elements of heavy-light light ray operators are used in a wide array of factorisation theorems for exclusive decays.
In this section we demonstrate a few important examples, where our results might be of use.
\subsection{B-meson distribution amplitudes}
The most prominent applications are without any doubt the distribution amplitudes of the B-meson. The two two-particle distribution amplitudes $\tilde{\phi}_B^+,\,\tilde{\phi}_B^-$ are defined by the following matrix element:
\begin{equation}
 \langle 0|\bar{q}_\beta(z)[z,0] (h_v)_\alpha(0)|B(p)\rangle =-i\frac{\hat{f}_B(\mu)}{4} 
  \left[(1+\fmslash{v})\left(\tilde{\phi}_B^+(t)+\frac{\fmslash{z}}{2t}[\tilde\phi_B^-(t)-\tilde\phi_B^+(t)] \right)\gamma_5\right]_{\alpha\beta},
\end{equation}
with $t=v\cdot z$ and $z^2=0$.
Inserting $\fmslash{n}$ or $\tilde{\fmslash{n}}$ one can project on $\tilde{\phi}_B^+$ or $\tilde{\phi}_B^-$, respectively. The relevant operators in spinor notation are then
\begin{equation}
\mathcal{O}_+(0,z)=S(h^i_v\otimes\psi^j_+),\qquad \mathcal{O}_-(0,z)=S(h^i_v\otimes\psi^j_-),
\end{equation}
where $S=\delta_{ij}$. By multiplying equation (\ref{eq:kern3}) with $\delta_{ij}$, 
setting $z_1=0$, $z_2=z$, and correcting for the renormalisation of the B-meson decay constant in HQET one recovers 
for $\phi_B^+$ the result of \cite{Braun:2003wx,Kawamura:2010tj}. After a Fourier-transformation the result of \cite{Lange:2003ff} follows.
The same can be done for $\phi_B^-$ by using equations (\ref{eq:kern4}) and (\ref{eq:kern5}). A Fourier-transformation confirms the 
results of \cite{Bell:2008er,DescotesGenon:2009hk}. Some care has to be taken in Fourier-transforming the $\log$-term:
\begin{align}
&\dfrac{1}{2\pi}\int_{-\infty}^{+\infty}dte^{i\omega t} \log (i\mu (t-i\epsilon))\mathcal{O}_\pm(0,t)\nonumber\\
&=\dfrac{i}{2\pi}\int_0^\infty d\omega'\left\{\dfrac{1}{\omega'-\omega-i\epsilon}\log\dfrac{\omega'-\omega-i\epsilon}{\mu}\right.\nonumber\\
&\left.-\dfrac{1}{\omega'-\omega+i\epsilon}\log\dfrac{\omega'-\omega+i\epsilon}{\mu}\right\}\mathcal{O}_\pm(\omega',\mu)\nonumber\\
&=-\log\dfrac{\mu}{\omega}\mathcal{O}_\pm(\omega,\mu)-\int_0^\infty d\omega'\left(\dfrac{\Theta(\omega'-\omega)}{\omega'-\omega}\right)_+\mathcal{O}_\pm(\omega',\mu),
\end{align}
where we define the $+$-distribution in the usual way
\[\int d\omega' \left(f(\omega,\omega')\right)_+\,g(\omega')=\int d\omega'f(\omega,\omega')g(\omega')-g(\omega)\int d\rho f(\rho,\omega).\] 
All other terms do not need any special treatment.
In principle, taking our results, it does not pose a problem to construct the renormalisation of many-particle distribution 
amplitudes with an arbitrary amount of quarks and transverse gluons but here we just want to comment on two 
three-particle distribution amplitudes. They were defined in \cite{Kawamura:2001jm} via the following matrix element: 
(for a more general approach see \cite{Geyer:2005fb, Geyer:2007yh})
\begin{eqnarray}
&&\langle 0|\bar{q}_\beta(z)[z,uz] gG_{\mu\nu}(uz)z^\nu [uz,0]
(h_v)_\alpha(0)|B(p)\rangle \nonumber\\
&&=\frac{\hat{f}_B(\mu) M}{4}
  \left[(1+\fmslash{v})
  \left[(v_\mu \fmslash{z} -
    t\gamma_\mu)\left(\tilde\Psi_A(t,u)-\tilde\Psi_V(t,u)\right)
    -i\sigma_{\mu\nu}z^\nu \tilde\Psi_V(t,u)\right.\right. \nonumber\\
&&\left.\left.\qquad\qquad\qquad\qquad
 - z_\mu \tilde{X}_A(t,u)
    +\frac{z_\mu \fmslash{z}}{t} \tilde{Y}_A(t,u)
   \right]\gamma_5\right]_{\alpha\beta}.
\end{eqnarray}
The combinations $\tilde{\Psi}_A-\tilde{\Psi}_V$ and $\tilde{\Psi}_A+\tilde{\Psi}_V$ can be identified as
\begin{equation}
 \mathcal{O}_{A-V}(0,uz,z)=S(h^i_v\otimes f^a_{++}\otimes \bar{\psi}^j_+),\qquad\mathcal{O}_{A+V}(0,uz,z)=S(h^i_v\otimes \bar{f}^a_{++}\otimes \bar{\psi}^j_-),
\end{equation}
with $S=t_{ij}^a$, while $\tilde{\Psi}_A$ and $\tilde{\Psi}_V$ are given as sum or difference of these operators and therefore 
cannot be associated with a well defined twist for the light degrees of freedom. Using the results from \cite{Braun:2009vc} and 
equations (\ref{eq:kern1}), (\ref{eq:kern3}) it is easy to construct the renormalisation of this distribution amplitudes. 
We confirm the results of \cite{Offen:2009mt} for $\Psi_A-\Psi_V$ after a Fourier-transformation.
\subsection{$\Lambda_b$ distribution amplitudes}
The $\Lambda_b$ distribution amplitudes are defined as matrix elements of non-local light ray operators
built of an effective heavy quark and two light quarks, see e.g. \cite{Ball:2008fw}:
\begin{eqnarray}
 \epsilon^{ijk}\langle 0\vert \left(u^{T\,i}(t_1 n)C\gamma_5\fmslash{n} d^j(t_2 n)\right)h_v^k(0)\vert \Lambda(v)\rangle&=&f_\Lambda^{(2)} \Psi_2(t_1,t_2)\Lambda(v),\nonumber\\
\epsilon^{ijk}\langle 0\vert \left(u^{T\,i}(t_1 n)C\gamma_5 d^j(t_2 n)\right)h_v^k(0)\vert \Lambda(v)\rangle&=&f_\Lambda^{(1)} \Psi^s_3(t_1,t_2)\Lambda(v),\nonumber\\
\epsilon^{ijk}\langle 0\vert \left(u^{T\,i}(t_1 n)C\gamma_5 i\sigma_{\tilde{n}n} d^j(t_2 n)\right)h_v^k(0)\vert \Lambda(v)\rangle&=&f_\Lambda^{(1)} \Psi^\sigma_3(t_1,t_2)\Lambda(v),\nonumber\\
\epsilon^{ijk}\langle 0\vert \left(u^{T\,i}(t_1 n)C\gamma_5\tilde{\fmslash{n}} d^j(t_2 n)\right)h_v^k(0)\vert \Lambda(v)\rangle&=&f_\Lambda^{(2)} \Psi_4(t_1,t_2)\Lambda(v).
\end{eqnarray}
$\Lambda(v)$ is a Dirac-spinor fulfilling $\fmslash{v}\Lambda(v)=\Lambda(v)$ where non-relativistic normalisation
$\bar{\Lambda}\Lambda=1$ is assumed, $\sigma_{\tilde{n}n}=\sigma_{\mu\nu}\tilde{n}^\mu n^\nu$, $C$ is the charge conjugation matrix which in spinor representation looks like
\begin{equation}
 C=\left(\begin{array}{cc} -\epsilon_{\alpha\beta}&0\\0&-\epsilon_{\dot{\alpha}\dot{\beta}}\end{array}\right)
\end{equation}
and the subscripts refer to the twist of the light diquark operator. $\Psi_3^\sigma$ is antisymmetric under interchange of the light 
quark coordinates while the others are symmetric. In spinor notation the relevant operators are
\begin{eqnarray}
 \mathcal{O}_2(t_1,t_2)&=&\epsilon^{ijk} \left(\psi_+^i(t_1)\, \bar{\chi}_+^j(t_2)\, h_v^k(0)\right),\nonumber\\
\mathcal{O}_3(t_1,t_2)&=&\epsilon^{ijk} \left(\psi_+^i(t_1)\, \psi_-^j(t_2)\, h_v^k(0)\right),\nonumber\\
\mathcal{O}_4(t_1,t_2)&=&\epsilon^{ijk} \left(\psi_-^i(t_1)\, \bar{\chi}_-^j(t_2)\, h_v^k(0)\right).
\end{eqnarray}
Using the results (\ref{eq:kern3}) for $h_v\otimes \psi_+$ and $\psi_+\otimes\bar{\psi}_+$ from \cite{Braun:2009vc} 
and correcting for the renormalisation of $f_\Lambda^{(2)}$ we recover the expressions of \cite{Ball:2008fw} but we 
can extend their result to the $\Psi_3$ case by using the necessary expressions for
\[h_v\otimes\psi_-,\quad h_v\otimes \psi_+,\quad\psi_+\otimes\psi_-,\]
from equations (\ref{eq:kern4}), (\ref{eq:kern5}) and from \cite{Braun:2009vc}. We will give a short outline of the calculation as an 
example of possible applications. The relevant kernel from \cite{Braun:2009vc} is
\begin{eqnarray}
 \left[\mathbb{H}\, \psi_+^i\,\psi_-^j\right](z_1,z_2)&=&-2 t^b_{ii'}t^b_{jj'}\left[\left[\hat{\mathcal{H}}\,\psi_+^{i'}\psi_-^{j'}\right](z_1,z_2)+2\sigma_q \psi_+^{i'}(z_1)\psi_-^{j'}(z_2)\right.\nonumber\\
&+&\left.\left[\mathcal{H}_{21}^{e,1}\,\psi_-^{i'}\psi_+^{j'}\right](z_1,z_2) \right]
\end{eqnarray}
with
\begin{eqnarray}
 \left[\hat{\mathcal{H}}\varphi\right](z_1,z_2)&=&\int_0^1d\alpha\left(2\varphi(z_1,z_2)-\bar{\alpha}^{2j_1-1}\varphi(z_{12}^\alpha,z_2)-\bar{\alpha}^{2j_2-1}\varphi(z_1,z_{21}^\alpha)\right),\\
\left[\mathcal{H}_{21}^{e,k}\varphi\right](z_1,z_2)&=&\int_0^1d\alpha\, \alpha^{2j_1-k-1}\alpha^{k-1}\varphi(z_{12}^\alpha,z_2).
\end{eqnarray}
After some simple colour algebra one can add up all necessary expressions resulting in the $3\to 3$ evolution of $\mathcal{O}_3$:
\begin{eqnarray}
 \left[\mathbb{H}^{(3\to3)}\mathcal{O}_3\right]&=&\dfrac{\alpha_s}{2\pi}C_F\epsilon^{ijk}\left[\int_0^1d\alpha\left(2\psi^i_+(z_1)\psi^j_-(z_2)-\bar{\alpha}\psi^i_+(z_{12}^\alpha)\psi^j_-(z_2)-\psi^i_+(z_1)\psi^j_-(z_{21}^\alpha)\right)h_v^k(0)\right.\nonumber\\
&+&\psi_+^i(z_1)\left\{\int_0^1d\alpha\left(\psi_-^j(z_2) h_v^k(0)-\psi_-^j(\bar{\alpha}z_2)h_v^k(0)\right)+\log(i\mu\,z_2)\psi_-^j(z_2)h_v^k(0)\right\}\nonumber\\
&+&\int_0^1d\alpha\left(\psi^i_+(z_1)\psi_-^j(z_2)h_v^k(0)-\bar{\alpha}\psi_+^i(\bar{\alpha}z_1)\psi_-^j(z_2)h_v^k(0)\right)-4\psi_+^i(z_1)\psi_-^j(z_2)h_v^k(0)\nonumber\\
&+&\log(i\mu\,z_1)\psi_+^i(z_1)\psi_-^j(z_2)h_v^k(0)+\left.\int_0^1d\alpha\, \psi_-^i(z_{21}^\alpha)\psi_+^j(z_1)h_v^k(0)\right].
\end{eqnarray}
The mixing with four particle operators is in this case completely governed by the kernels (\ref{eq:kern5}) and comparatively short:
\begin{eqnarray}
 \left[\mathbb{H}^{(3\to4)}\mathcal{O}_3\right]&=&\dfrac{\alpha_s}{2\pi}\epsilon^{ijk}\left[\psi_+^i(z_1)\left\{f^{abc}t^b_{jj'}t^c_{kk'}z_2^2\int_0^1 d\alpha\int_0^{\bar{\alpha}}d\beta\,\beta\, \psi_+^{j'}(\bar{\alpha}z_2)\bar{f}_{++}^a(\beta z_2)\right.\right.\nonumber\\
&+&\left.i(t^at^b)_{jj'}t^b_{kk'}z_2^2\int_0^1d\alpha\int_{\bar{\alpha}}^1d\beta\dfrac{\bar{\alpha}\bar{\beta}}{\alpha}\psi_+^{j'}(\bar{\alpha}z_2)\bar{f}_{++}^a(\beta z_2)\right\}h_v^{k'}(0)\nonumber\\
&+&\psi_+^{i'}(z_1)\left\{f^{abc}t^b_{ii'}t^c_{jj'}z_{12}^2\int_0^1 d\alpha\int_0^{\bar{\alpha}}d\beta\,\beta\,\psi_+^{j'}(z_{21}^\alpha)\bar{f}_{++}^a(z_{12}^\beta)\right.\nonumber\\
&+&\left.\left.i(t^a t^b)_{jj'}t^b_{ii'}z_{12}^2\int_0^1d\alpha\int_{\bar{\alpha}}^1d\beta\dfrac{\bar{\alpha}\bar{\beta}}{\alpha}\psi_+^{j'}(z_{12}^\alpha)\bar{f}_{++}^a(z_{21}^\beta)\right\}h_v^k(0)\right].
\end{eqnarray}
As one can see, the pattern is similar as in the twist 3 pseudoscalar meson case. There the operators
\[\mathcal{O}^1_3(z_1,z_2)=\chi_+(z_1)\psi_-(z_2),\quad \mathcal{O}^2_3(z_1,z_2)=\chi_-(z_1)\psi_+(z_2),\quad\mathcal{O}^3_3(z_1,z_2,z_3)=\chi_+(z_1)\bar{f}_{++}(z_2)\psi_+(z_3)\]
build a closed set under renormalisation. Here we have
\begin{eqnarray}
 \mathcal{O}^1_3(t_1,t_2)&=& \psi_+(t_1)\, \psi_-(t_2)\, h_v(0),\nonumber\\
 \mathcal{O}^2_3(t_1,t_2)&=& \psi_-(t_1)\, \psi_+(t_2)\, h_v(0),\nonumber\\
\mathcal{O}^3_3(t_1,t_2,t_3)&=& \psi_+(t_1)\, \psi_+(t_2)\,\bar{f}_{++}(t_3) h_v(0).
\end{eqnarray}

%Using the kernels for $h_v\otimes \psi_+$ and $\psi_+\otimes \psi_+$ and making a Fourier-transformation it is relatively easy to reproduce the results from \cite{Ball:2008fw}.
%\subsection{Subleading shapefunctions in $\bar{B}\to X_s\gamma$-decays}
%As a last example lets take a look at the decay $\bar{B}\to X_s\gamma$. In a recent study \cite{Benzke:2010js, Benzke:2010tq} effects
%of subleading shapefunctions entering the factorisation theorem at order $\frac{1}{m_b}$ were thoroughly analyzed.  
%EVENTUELL RAUSSTREICHEN, DA NICHT SICHER IST, OB MAN UNSERE ERGEBNISSE DORT WIRKLICH GEBRAUCHEN KANN!
\section{Conclusions and summary}
We have calculated the renormalisation of four different heavy-light light ray operators. Besides confirming results of \cite{Lange:2003ff, Braun:2003wx, Bell:2008er, DescotesGenon:2009hk, Offen:2009mt, Kawamura:2010tj}
we were able to show that all $2\to2$-kernels are given by a single function (\ref{eq:2kernels}) and that this function is 
determined up to a constant by the pattern of conformal symmetry breaking (\ref{eq:invar1a}), (\ref{eq:invar1b}) and (\ref{eq:invar3}). 
Furthermore one could in principle go the other way round where one only has to calculate the divergent part of the insertion of 
the conformal anomaly into an one loop Greens-function of the relevant operator. Using that the resulting commutation relations (\ref{eq:counterterm-form})
are a general feature of the renormalisation of heavy-light light ray operators one can then construct all the $2\to2$-renormalisation kernels except for an unknown constant.\\
The breaking of conformal symmetry already for the one loop counterterms can be traced back to the cusp in 
two Wilson-lines one light-like required for the gauge-invariance of the operators and one time-like representing 
the effective heavy quark field. This cusp in the path of the Wilson-lines requires an additional UV-renormalisation 
given by $\Gamma_{cusp}$ and leads to the difference to full QCD, namely that the insertion of the conformal anomaly in the 
Greens-function of a heavy-light light ray operator gives already at one loop a divergent piece which prevents that the one loop
counterterms exhibit conformal symmetry. As noted in section 4 this statement is only valid for the two-particle counterterms while 
the three-particle terms stay free of these additional symmetry-breaking divergences. A fact not exploited further but it hints towards 
a justification why, for the cases at hand, the $2\to3$ mixing is governed solely by the twist of the light degrees of 
freedom. We have shown by explicit calculation that the $2\to3$-kernels of $h_v\otimes\psi_-$ and $h_v\otimes f_{+-}$ coincide with 
those of $\psi_+\otimes\psi_-$ and $\psi_+\otimes f_{+-}$, respectively and could not find any additional mixing due to the heavy quark.\\
Our results can be seen as a first step towards a systematic analysis of the renormalisation of heavy-light light ray operators and 
they enable us to construct the renormalisation of several leading and non-leading distribution amplitudes of heavy-light 
mesons or baryons. In principle it is even possible to include an arbitrary number of gluons by using the kernels (\ref{eq:kern1}) and 
(\ref{eq:kern2}).
\section*{Acknowledgements}
We are grateful to V.M. Braun, S. Descotes-Genon and A. Manashov for numerous discussions and helpful hints concerning our results. 

\begin{appendix}
\section{$Z_h$ in light cone gauge}
\begin{figure}[h]
 \begin{center}
  \epsfig{file=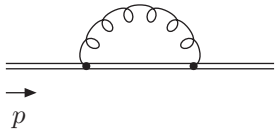, scale=1.0}
\caption{Renormalisation of the heavy quark. \label{fig:ren-heavy}}
 \end{center}
\end{figure}
Here we calculate the diagram shown in Fig. \ref{fig:ren-heavy} in light cone gauge. 
We use an off-shell momentum and a gluon mass as infrared regulators to extract solely the UV-divergences. We again use 
the Mandelstam-Leibbrandt \cite{Mandelstam:1982cb, Leibbrandt:1983pj} prescription (\ref{eq:Mand-Leib}) for the extra pole 
in the gluon propagator. The resulting expression is
\begin{equation}
 -C_F g_s^2\left(\dfrac{\mu}{2\pi}\right)^{4-D}\int \dfrac{d^Dl}{(2\pi)^D}\left[\dfrac{1}{v\cdot(p+l)(l^2-m^2)}-2\dfrac{v\cdot l}{n\cdot l\, v\cdot(p+l) (l^2-m^2)}\right],
\end{equation}
where one clearly sees the contribution equivalent to the Feynman-gauge and the additional term due to the modification of the 
gluon propagator. The second term can be rewritten as
\begin{equation}
 2C_f g_s^2\left(\dfrac{\mu}{2\pi}\right)^{4-D}\int \dfrac{d^Dl}{(2\pi)^D}\left(\dfrac{1}{n\cdot l(l^2-m^2)}-\dfrac{v\cdot p}{n\cdot l\, v\cdot(p+l)(l^2-m^2)}\right).
\label{eq:heavy-add-part}
\end{equation}
The first term vanishes, since the poles of the two denominators always lie in the same half plane, while the second one matches 
the first integral in the third row of equation (\ref{eq:res-calc1}) except for the regulators. Taking the integrals gives the 
following expression for the renormalisation constant of the heavy quark in light cone gauge: 
\begin{equation}
 Z_h^{lc}=1+\dfrac{\alpha_s}{2\pi}C_F\dfrac{1}{\varepsilon}+\dfrac{\alpha_s}{2\pi}C_F\left[\dfrac{1}{\varepsilon^2}-\dfrac{1}{\varepsilon}\log\dfrac{(2v\cdot p)^2}{\mu^2}\right].
\end{equation}
Since the second term in (\ref{eq:heavy-add-part}) cancels in gauge invariant operators always against mentioned integral in (\ref{eq:res-calc1})
we only need the Feynman-gauge result and therefore define $\sigma_h$ as:
\begin{equation}
 Z_h^{1/2}=1+\dfrac{\alpha_s}{2\pi}C_F\dfrac{1}{\varepsilon}\sigma_h,\qquad\sigma_h=\dfrac{1}{2}.
\label{eq:sigmah}
\end{equation}

%\section{Conformal transformations}
\section{Variation of the action under dilatation and special conformal transformation in Feynman- and light cone gauge}
\begin{figure}[h]
\begin{center}
 \epsfig{file=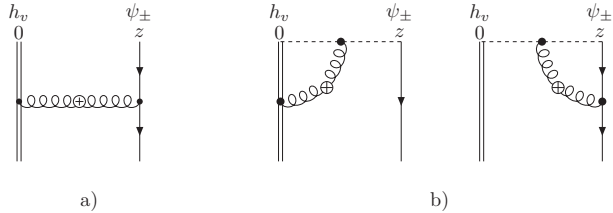, scale=0.75}
\caption{Diagrams for heavy quark anti-quark operator which are needed to show that the insertion of the operators 
(\ref{eq:BRST-op-lc}) and (\ref{eq:BRST-op-fg})  does not give a contribution.\label{Fig:BRST}}
\end{center}
\end{figure}

Here we give some details concerning the variation of the action under dilatation and special conformal transformation and we show 
that the additional operators which were not considered in section 4 give no contributions for gauge invariant operators.\\
In a covariant gauge the variation of the action and gauge fixing terms under dilatation and special conformal transformation takes 
the following form \cite{Belitsky:1998vj, Belitsky:1998gc, Braun:2003rp}:
\begin{eqnarray}
 \delta_D\,S&=&\varepsilon \int d^Dx\left[\mathcal{O}_A+\mathcal{O}_B+\Omega_{\bar{\omega}}-\Omega_{\bar{\psi}\psi}\right],\\
\delta_K^\nu\,S&=&2\varepsilon \int d^Dx\,x^\nu\left[\mathcal{O}_A+\mathcal{O}_B+\Omega_{\bar{\omega}}-\Omega_{\bar{\psi}\psi}\right]\nonumber\\
&+&2(D-2)\int d^Dx\,x^\nu\partial_\mu\mathcal{O}_{B}^\mu,\qquad\varepsilon=\dfrac{1}{2}(4-D).
\end{eqnarray}
In light cone gauge the violation of Lorentz-symmetry and scale invariance makes the result slightly more complicated:
\begin{eqnarray}
 \delta_D\,S&=&\varepsilon \int d^Dx\left[\mathcal{O}_A+\mathcal{O}_B+\Omega_{\bar{\omega}}-\Omega_{\bar{\psi}\psi}\right]\nonumber\\
&-&\int d^Dx\,\mathcal{O}_B,\label{eq:var-dil-lc}\\
\delta_K^\nu\,S&=&2\varepsilon \int d^Dx\,x^\nu\left[\mathcal{O}_A+\mathcal{O}_B+\Omega_{\bar{\omega}}-\Omega_{\bar{\psi}\psi}\right]\nonumber\\
&-&2\int d^Dx\,x^\nu\mathcal{O}_B-2\int d^Dx\,x^\nu\partial_\rho\left( n^\rho x^\mu-n\cdot x g^{\rho\mu}\right)\mathcal{O}_{B\mu}.\label{eq:var-vk-lc}
\end{eqnarray}
We use the notation of \cite{Belitsky:1998vj, Belitsky:1998gc, Braun:2003rp} for the different appearing operators
\begin{equation}
 \begin{array}{lll}
\mathcal{O}_A(x)=\dfrac{1}{2} \left(G_{\mu\nu}^a\right)^2,& 
\mathcal{O}_B(x)=\dfrac{\delta^{BRST}}{\delta \lambda}\bar{\omega}^a\partial^\mu A_\mu^a,&
\mathcal{O}_{B\mu}(x)=\dfrac{\delta^{BRST}}{\delta \lambda}\bar{\omega}^a\,A_\mu^a,\\&&\\
\Omega_A(x)=A^a_\mu\dfrac{\delta S}{\delta A_\mu^a},&
\Omega_{\bar{\psi}\psi}(x)=\dfrac{\delta S}{\delta \psi}\psi+\bar{\psi}\dfrac{\delta S}{\delta \bar{\psi}},&
\Omega_{\bar{\omega}}(x)=\bar{\omega}^a\dfrac{\delta S}{\delta \bar{\omega}^a} 
\end{array}
\end{equation}
where $\omega$ and $\bar{\omega}$ are ghost and anti-ghost fields, respectively.
In light cone gauge one has to substitute $\partial^\mu$ by $n^\mu$ in $\mathcal{O}_B$ and 
one has to be aware that the BRST-transformations differ in covariant and axial gauges.
Of special interest for our argument in section 4 are those operators $\mathcal{O}_B,\,\mathcal{O}_{B\mu}$ which do not come with an 
$\varepsilon$-factor and which include two gluon fields due to
\begin{eqnarray}
 \delta^{BRST} \bar{\omega}^a&=&\dfrac{1}{\xi}\partial^\mu A_\mu^a\,\delta\lambda\qquad \mbox{in covariant gauge,}\\
\delta^{BRST} \bar{\omega}^a&=&\dfrac{1}{\xi}n^\mu A_\mu^a\,\delta\lambda\qquad \mbox{in axial gauge,}
\end{eqnarray}
where $\delta\lambda$ is an anticommuting Grassmann-number.\footnote{The operator $\int d^Dx\, \mathcal{O}_B$ appearing in (\ref{eq:var-dil-lc}) and 
(\ref{eq:var-vk-lc}) does not give a contribution upon insertion since both gluon propagators are contracted with $n$ and 
therefore the result is proportional to $\xi$ which vanishes in light cone gauge. See below.}
It is now a straightforward task to show that the insertion of the resulting operators, here we show only the relevant gluon-field 
part, 
\begin{eqnarray}
\mathcal{O}_{v\cdot K}^A=\dfrac{2}{\xi}\int d^Dx\left(x^\mu-n\cdot x v^\mu\right)A^a_\mu n^\nu A^a_\nu(x)\qquad \mbox{in axial gauge,}
\label{eq:BRST-op-lc}\\  
\mathcal{O}_{v\cdot K}^A=\dfrac{2}{\xi}(D-2)\int d^Dx\,v^\mu\,A^a_\mu\partial^\nu A^a_\nu(x)\qquad \mbox{in covariant gauge,}\label{eq:BRST-op-fg}
 \end{eqnarray}
vanishes. Figure \ref{Fig:BRST} shows the necessary diagrams which have to be calculated for the simplest case of a 
gauge invariant heavy quark, anti-quark operator.
In light cone gauge the only relevant diagram is always the exchange diagram and it vanishes for all two-particle operators, 
the only subtlety being, that since the operators are proportional to $\frac{1}{\xi}$ one has to take into account contractions 
of the gluon propagator with $n^\mu$ which are proportional to $\xi$:
\begin{eqnarray}
 n^\mu d_{\mu\nu}^{ab}(q,\xi)&=&-in^\mu\dfrac{\delta^{ab}}{q^2+i\epsilon}\left(g_{\mu\nu}-\dfrac{q_\mu n_\nu+q_\nu n_\mu}{n\cdot q}+q_\mu q_\nu\dfrac{n^2+\xi q^2}{(n\cdot q)^2}\right)\nonumber\\
&=&-i\xi\dfrac{q_\nu}{n\cdot q}.
\end{eqnarray}
In covariant gauges the contributions from insertion of $\mathcal{O}_{v\cdot K}^A$
only vanish if one considers gauge invariant heavy-light light ray operators 
as in figure \ref{Fig:BRST} b). This can be understood in the following way: Since in covariant gauges the $\log$- and 
integral-term in eq. (\ref{eq:2kernels}) would have different colour structures the constraint (\ref{eq:invar1b}) 
would be proportional to the difference of these. The insertion of $\mathcal{O}_A$ gives only a finite result as 
seen in (\ref{eq:invar_K}) and therefore does not explain this result.  Only the insertions of $\mathcal{O}_{v\cdot K}^A$ 
give the divergences with exactly the right colour structures to account for the changed commutator relation (\ref{eq:invar1b}).

%In light cone gauge this is true for all two-particle operators while in covariant gauges one has to consider gauge invariant 
%operators where the divergences in different diagrams cancel as for example in figure \ref{Fig:BRST} b).

%Both can be written in the following unified way \cite{Balitsky:1987bk, Braun:2003rp}
%\[\]
\end{appendix}

\end{document}